\documentclass[12pt,preprint]{aastex}

\newcommand\teff{T_{\rm eff}}

\newcommand{\wig}[1]{\mathrel{\hbox{\hbox to 0pt{%
          \lower.6ex\hbox{$\sim$}\hss}\raise.4ex\hbox{$#1$}}}}

\shorttitle{short title}
\shortauthors{Saumon et al.}

\begin{document}


\title{Near-UV absorption in very cool DA white dwarfs}


\author{D. Saumon}
\affil{Los Alamos National Laboratory, PO Box 1663, Mail Stop F663, Los Alamos, NM 87545}
\email{dsaumon@lanl.gov}

\author{J. B. Holberg}
\affil{Lunar and Planetary Laboratory, University of Arizona, Sonett Space Sciences Building, 
       1541 East University Boulevard, Tucson, AZ 85721-0063}
\email{holberg@argus.lpl.arizona.edu}

\author{P. M. Kowalski}
\affil{Institute of Energy and Climate Research (IEK-6), Forschungszentrum J\" ulich,
       Wilhelm-Johnen-Strasse, 52425 J\"ulich, Germany}
\email{p.kowalski@fz-juelich.de}



\begin{abstract}
The atmospheres of very cool, hydrogen-rich white dwarfs ($\teff <6000\,$K)
are challenging to model because of the increased complexity of the equation of state,
chemical equilibrium, and opacity sources in a low-temperature, weakly ionized
dense gas.  In particular, many models that assume relatively simple models for the
broadening of atomic levels and mostly ideal gas physics overestimate the flux
in the blue part of their spectra.  A solution to this problem that has met with some success
is that additional opacity at short wavelengths comes for
the extreme broadening of the Lyman $\alpha$ line of atomic H by collisions
primarily with H$_2$.  For the purpose of validating this model more rigorously, we
acquired {\it Hubble Space Telescope} STIS spectra  of eight very cool white dwarfs (5 DA and 3 DC stars).
Combined with their known parallaxes, $BVRIJHK$ and {\it Spitzer} IRAC photometry, we analyze
their entire spectral energy distribution (from 0.24 to 9.3$\,\mu$m) with a large
grid of model atmospheres and synthetic spectra.  We find that the red wing of the
Lyman $\alpha$ line reproduces the rapidly decreasing near-UV flux of these very
cool stars very well. We determine better constrained values of $\teff$ and gravity as 
well as upper limits to the helium abundance in their atmospheres.

\end{abstract}





\section{Introduction}

As the end stage of the evolution of the vast majority of stars, white dwarfs (WDs) are 
common denizens of the solar neighborhood. With masses of about 0.6$\,$M$_\odot$ and
radii of $\sim 10^9\,$cm, they are characterized by high surface gravities
of $\log g \,({\rm cm/s}^2) \sim 8$. The coolest known disk white dwarfs have effective
temperatures of $\teff \sim 4000\,$K or slightly lower \citep{blr01,kilic09a,kilic10a,giam12}. The combination of low temperature and high gravity 
results in high photospheric densities where dense matter physics comes into play and affects the
model atmospheres and emergent spectra significantly.
The atmospheres of very cool white dwarfs present modeling challenges that are not 
encountered in other types of stars.

White dwarf atmospheres thus present interesting physics problems in the equation of state of
non-ideal gases, molecular dissociation and ionization in the presence of inter-atomic 
and inter-molecular interactions, extreme pressure-broadening of absorption lines, and other
dense matter physics which are unique in the field of stellar atmospheres \citep{kowalski10}.
Laboratory opacity measurements have started to probe this interesting regime 
only recently \citep{falcon13}. On the other hand, the spectra of these very cool white dwarfs 
bear the signatures of dense matter physics and models of these effects can thus be tested in a fairly direct fashion.

In addition to their value as good laboratories for the physics of dense hydrogen and helium,
very cool white dwarfs are of interest as common remnants of ordinary 
stars.  Devoid of internal energy sources, they simply cool to ever lower effective temperatures. The
finite age of the Galaxy results in a fairly sharp drop in the white dwarf luminosity function
at $\log L/L_\odot \sim -4.5$ \citep{harris06} or $\teff \sim 4000\,$K. 
The shape of the WD luminosity function -- the location of the low-luminosity drop off in particular -- contains
information about the age of the WD population, 
the star formation history of its progenitors and the physics of WD cooling as well \citep{fbb01}.  This
approach to cosmochronology has led to estimates of
the age of the disk of the Galaxy \citep{winget87,wood92,lrb98,giam12} and of several globular and galactic clusters 
\citep{hansen04,vonhippel05,hansen07,jeffery11,hansen13}. Such determinations depend primarily on the estimated 
ages of the coolest
WDs of the sample, requiring accurate values of their basic stellar parameters such as
$\teff$, gravity, and composition \citep{kowalski07}. For disk WDs the drop off in the luminosity function is
bright enough to be observed and  the accuracy of WD cosmochronology is determined in large
part by the ability of atmosphere models and synthetic spectra in matching the 
spectral energy distributions (SEDs) of the coolest known WDs.

In a landmark analysis of homogeneous optical and near-infrared photometry of a large sample of cool
white dwarfs, \citet{brl97} and \citet{blr01} found that their models systematically overestimated the flux in the $B$ band
of WDs of spectral type DA (showing Balmer lines in the spectra)  with $\teff \wig< 5300\,$K,
suggesting that a source of opacity was missing in the calculations at short wavelengths. 
This deficiency of the models at short wavelengths has sometimes forced several studies to exclude from the analysis the $B$ flux 
\citep{brl97,blr01} or the Sloan Digital Sky Survey (SDSS, \citet{sdss}) $u$ and sometimes $g$ fluxes \citep{kilic10a} for the coolest stars. 
\citet{brl97} suggested that the most likely source of this ``blue blanketing'' was a broadening of the Lyman bound-free edge 
of hydrogen atoms (at $911.8\,$\AA) to longer wavelengths due to a lowering of the continuum by collisions.
Modeling this ``pseudo-continuum'' opacity in an {\it ad hoc} fashion, they showed that
it could, in principle, reproduce the observed $B$ fluxes. 

An interesting illustration of this problem is the case of the cool DA star BPM 4729 (WD 0752$-$676). 
\citet{brl97} and \citet{blr01} found and excellent fit of the $VRIJHK$ photometry and the H$\alpha$ line
with a pure hydrogen atmosphere with $\teff=5730\,$K and $\log g=8.21$. \citet{wolff02} re-analyzed this star  
adding to the photometry a near-ultraviolet (near-UV) spectrum from the {\it Hubble Space Telescope} ({\it HST}) Faint Object Spectrograph that
extends the observed SED down to $\sim 0.25\,\mu$m. With models that include the pressure broadening of the 
Ly$\alpha$ line, they found that pure hydrogen models failed to reproduce the SED below $0.4\,\mu$m. Only by
considering atmospheres dominated by helium, with a number ratio of $N({\rm H})/N({\rm He})=3\times10^{-5}$ 
were they able to reproduce the near-UV spectrum, with $\teff=5500\,$K and $\log  g=8.21$. However, this
model did not provide a good match of the $IJHK$ photometry and, due to the trace amount of hydrogen, 
did not produce the H$\alpha$ and H$\beta$ lines seen in the spectrum. 

By considering the interactions between the H atoms and H$_2$ molecules that  
form in the atmosphere of a very cool DA, \citet{kowalski06b} provided a realistic  model of the bound-free
opacity and the broadening of the Ly$\alpha$ edge and showed that it is negligible in
the blue part of the spectrum of cool WDs. On the other hand, using a similar broadening model, \citet{ks06} 
showed that the Ly$\alpha$ line could be collision-broadened well into the blue part of the optical
spectrum. Collisions of neutral H atoms with H, He and H$_2$ were modeled and
\citet{ks06} showed that in very cool white dwarfs the broadening of Ly$\alpha$ was caused primarily by 
collisions with H$_2$ molecules.  Subsequently,
\citet{ak09} computed Ly$\alpha$ profiles broadened by collisions with H and He but not H$_2$.
As a result of this omission, their profiles do not extend into the optical and 
are unable to fully explain the flux deficiency observed in very cool WDs. The Ly$\alpha$ profile was 
independently modeled by \citet{rohrmann11}, including collisions with H$_2$, nicely confirming the 
calculations of \citet{ks06}.

When implemented in atmosphere models, the opacity of the red wing of the Ly$\alpha$ line
is able to reproduce the photometry
of the DA and DC stars in the BRL sample quite well, including the problematic $B$ band flux \citep{ks06}. 
It also matches the WD sequence in the $ugz$ color-magnitude diagram of hundreds of WDs in the SDSS
sample of \citet{kilic06}.  A more critical test of this new opacity model 
was a fit of the spectral energy distribution of BPM 4729 including the near-UV spectrum. \citet{ks06}
produced the first good fit of the entire SED of this star. In contrast with previous efforts, they found
that a pure hydrogen atmosphere model nicely reproduces its spectrum, with $\teff=5820\,$K and $\log g=8.30$.
This represented a good but limited validation of their Ly$\alpha$ opacity model.  More recent analyses of
WD photometry have successfully matched the blue part of the SEDs with these Ly$\alpha$ profiles
\citep{kilic09a,kilic09b,giam12}.  To summarize, the opacity provided
in the near-UV and blue part of the spectra of very cool WDs by the red wing of the Ly$\alpha$ line as modeled
by \citet{ks06} and \citet{rohrmann11} is on a sound theoretical basis. Empirically, it is well-supported by comparison with
fits of broad band photometry of dozens of white dwarfs, but these SEDs include only one bandpass 
that falls within the line wing ($B$ or $g$). The model passed a more stringent test by reproducing
not just a broadband flux ($B$) but also the {\it shape} of the profile for BPM 4729. 
Together, these successful applications of the opacity model for the red wing of Ly$\alpha$ constitute valuable but limited
tests. A more direct and thorough approach is desirable to firmly establish its validity.

In this contribution, we present new near-UV spectroscopy of eight very cool white dwarfs with $\teff < 6000\,$K 
acquired with the Space Telescope Imaging Spectrograph (STIS; \citet{stis}) on the {\it HST}. These 
data are combined with optical, near-IR and mid-IR photometry, and the 
stars' parallaxes to produce complete spectral energy distributions.   We fit these data with 
synthetic spectra with the goals of more thoroughly validating the model of
\citet{ks06} for the red wing of the Ly$\alpha$ line and determine more reliable $\teff$ and gravities
for our sample stars.
The possible presence of He in the atmosphere of cool DA and DC stars has long been of interest to understand 
the mechanism of spectral evolution of cool white dwarfs. The determination of the He abundance
remains fraught with uncertainty in the absence of He absorption lines in these stars.
As the {\it HST} STIS data extends the SEDs of our stars considerably, and that the broadening of the Ly$\alpha$ line
is sensitive to the presence of He, we also explore what constraints can be obtained on the He/H ratio in the 
atmospheres of very cool DA and DC white dwarfs.  

\section{Sample, observations and other data}

\subsection{Sample of very cool white dwarfs}

To achieve these goals, we selected eight very cool WDs, five of spectral type DA and three of type DC to be observed with
{\it HST}/STIS. The targets were required to have $\teff\wig< 6000\,$K so that broadening of the Ly$\alpha$ line by H$_2$
would be expected, to have H-dominated atmospheres as determined by recent studies, and to be bright enough
to be detectable with a sufficient signal-to-noise with the STIS instrument. Furthermore, and to maximally constrain
the models, each target was required to have a known trigonometric parallax, $BVRI$ optical photometry, 
Two Micron All Sky Survey (2MASS; \citet{2mass}) $JHK_s$ photometry, and {\it Spitzer Space Telescope} Infrared Array Camera 
(IRAC; \citet{irac}) fluxes in all four bands.  While we selected very cool stars to cover a broad 
range of $\teff$, we specifically excluded so-called ``ultracool'' WDs. These DC stars are
characterized by a very strong flux deficit in the near-IR and very peculiar SEDs \citep{gates04, harris08}
that cannot be reproduced by current models \citep{kilic09a,kilic10a}. These poorly understood stars are unsuitable for our purposes.

\subsection{{\it HST} STIS Observations}
  \label{stis_obs}

\begin{figure}
   \plotone{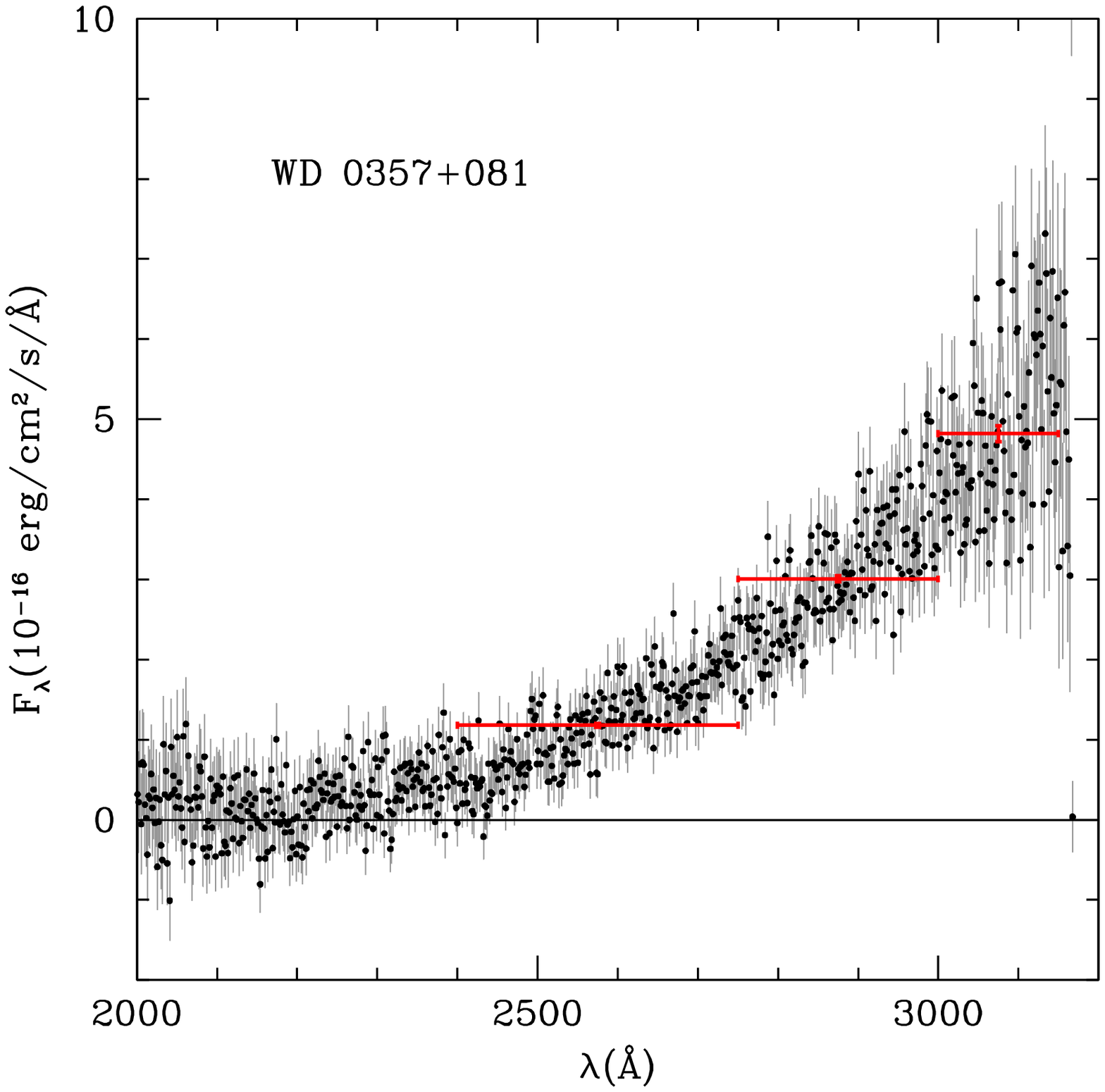}
   \caption{{\it HST} STIS spectrum of WD 0357+081. Horizontal bars show the box car bandpasses
            adopted to convert the STIS spectra into photometric data for the analysis
            (STIS1, STIS2, and STIS3, from left to right). The height of each bar gives the
            average flux and a vertical bar shows the $1\sigma$ uncertainty.
           [{\it See the electronic edition of the Journal for a color version of this figure.}]}
    \label{stis_sp}
\end{figure}

The STIS observations of our eight targets were conducted under {\it HST} Cycle 18 GO program 12188 
during 2011 and 2012.  The primary observing sequence in each instance was identical.  
The spectra were obtained with the G230L grating in first order and the $50\arcsec \times 2\arcsec$
aperture with the STIS/NUV-MAMA detector in ``accumulate'' mode.  The one-orbit exposures, 
which ranged from 2000 to 2990$\,$s, covered the wavelength range 1600 to $3175\,$\AA \  with 
a nominal resolving power of 500.  Because our targets are relatively faint and 
characterized by spectral energy distributions that steeply plunge towards shorter 
wavelengths, useful data were only obtained longward of 2400$\,$\AA.  
A typical STIS spectrum is shown in Figure \ref{stis_sp}.  All of the  STIS spectra are featureless within
the noise level of the data. 

For the purpose of fitting the SED of each star, we converted the STIS spectra into photometric fluxes
in three bandpasses. This effectively puts the STIS data on the same footing as the optical and infrared photometry
for the analysis.  To maintain a relatively uniform sampling of
the SED and weighting of the various spectral regions in the fitting procedure, we defined bandpasses with
$\lambda /\Delta \lambda \sim 10$, which is comparable to the resolving power of standard optical and 
near-IR filters. We also
widened the bandpasses at shorter wavelengths to maintain a reasonable S/N ratio as the flux decreases. The adopted
boxcar bandpasses to bin the STIS spectra are 2400 -- 2750$\,$\AA \ (``STIS1''), 2750 -- 3000$\,$\AA
\ (``STIS2''), and 3000 -- 3150$\,$\AA \ (``STIS3'') as shown in Figure \ref{stis_sp}.  The resulting 
fluxes and uncertainties are given in Table \ref{tab_stis_fluxes}. 

Depending on the band, detected fluxes were at the level of several times $10^{-18}\,{\rm erg}\,{\rm s}^{-1}\,{\rm cm}^{-2}{\rm \AA}^{-1}$
corresponding to S/N ratios of 6-50$\sigma$.  
However, no detectable signal was measured for our faintest target, WD 0747+073A.
Another faint target, WD 0747+073B, was barely detected in our STIS3 band at the 
$\sim 1\sigma$ level.  For these two stars only upper limits were obtained that are consistent 
with background limited uncertainties.  It appears likely that these two stars were not properly acquired
in the STIS slit and that the pointing missed both targets. WD 0747+073A is our faintest target but is only  about 30\% fainter 
in the optical than our second faintest target, WD 2054$-$050 (Table \ref{tab_opt_fluxes}). The other undetected star,
WD 0747+073B, is comparable in brightness to WD 2054$-$050 in the optical which was
cleanly detected in all three STIS bandpasses at $\ga 6\sigma$. Considering that the SEDs of our samples stars are all rapidly 
decreasing toward shorter wavelengths, we would expect that a somewhat faint object would show some flux in
the STIS3 and perhaps STIS2 bandpasses, but this is not the case here. These two undetected targets form a common proper
motion pair separated by $16\arcsec$. WD 0747+073A was acquired by offset from the position of its companion. Thus,
a pointing error in the STIS observations of WD 0747+073B would also compromise the acquisition of WD 0747+074A.  
Further evidence that only blank sky was observed for these two stars is provided by our analysis of their SEDs
(\S \ref{results}).  In light of these considerations, we exclude the upper limits on the STIS fluxes given 
in Table \ref{tab_stis_fluxes} from our analysis of these two stars.

\subsection{Photometry and fluxes}

Optical photometry in the Landolt $BVRI$ bandpasses was obtained primarily from \citet{brl97}, complemented with
data from additional published sources.  In general averages were taken when there was more than one set of data for a star.
$U$ band photometry is desirable as it bridges the ground-based optical data and the near-UV STIS data.
It is obtained from the ground and is well-known
to be affected by the strong and steeply increasing atmospheric absorption towards shorter wavelengths. In practice,
it is subject to large systematic errors that are not necessarily reflected in the quoted measurement uncertainties. 
Five of our targets have published $U$ magnitudes.
For WD 1820+609 the $U$ flux is strongly discrepant with the neighboring STIS and $B$
fluxes and is thus suspicious.  Of the remaining four stars, our fits of the SEDs (see below) reveal the $U$
flux to be very poorly matched in three cases, even though the adjacent fluxes are well fitted. Only
for WD 2054$-$050 were we able to obtain a good fit of the $U$ flux.   For these reasons, we consider the
$U$ magnitudes of these stars to be unreliable and have ignored them in the analysis.

Near-infrared $JHK_s$ photometry for our targets 
was drawn from the 2MASS Point Source Catalog\footnote{http://www.ipac.caltech.edu/2mass/releases/allsky/}. The 
$BVRIJHK_s$ magnitudes were converted to flux units following the calibration of \citet{hb06}. Mid-infrared
fluxes are available in all four channels of the {\it Spitzer} IRAC instrument \citep{kilic09a}. 
Tables \ref{tab_stis_fluxes}--\ref{tab_ir_fluxes}  compile all the fluxes as well as the parallaxes used in our analysis.
The combination of the three STIS fluxes, $BVRIJHK_s$ fluxes and all four IRAC fluxes provides an evenly spaced sampling of the
SEDs of our targets from 0.24 to 9.3$\,\mu$m with only one gap at the $U$ bandpass.  In $F_\nu$ units, the flux distributions
typically peak in the $I$ or $J$ bandpass and reach down to fluxes 100 times fainter at the blue end of our data, and 
$\sim 10$ times fainter at the red end. Additionally, H$\alpha$ spectroscopy is available for all targets \citep{giam12}.
This represents a remarkably complete data set for such very cool WDs.

\section{Analysis}

\subsection{Atmosphere models}
\label{models}

The equation of state, the chemical equilibrium and the opacity of the gas are all affected by
the high densities that occur in the atmospheres of very cool white dwarfs. 
The atmosphere models and synthetic spectra  we applied in the following analysis include all the dense fluid models 
we developed for WD atmospheres over the last decade.  Specifically, they include for hydrogen
the pressure-broadened red wing of the hydrogen Ly$\alpha$ line \citep{ks06} -- the focus of this study --,
the non-ideal dissociation equilibrium of $\rm H_2$ \citep{kowalski06a}, and the pressure induced ionization of $\rm H^-$ \citep{kowalski10}.
For helium, density corrections to the Rayleigh scattering, the $\rm He^-$ free-free opacity \citep{irs02}, and 
the non-ideal ionization equilibrium in the low-temperature, dense fluid \citep{kowalski07_he} are all important for He-rich models
and especially pure He models.  In mixed He/H atmospheres, we used $\rm H_2-He$ collision-induced absorption recently calculated by \citet{abel12}.
Finally, the solution of the radiative transfer equation takes into account the vertical variation of the refractive index \citep{ks04}. 
Because the gas density at the photosphere remains moderate for He/H ratios of $y \equiv \log N({\rm He})/N({\rm H}) \le 2$, we consider 
the models of low He/H ratios to be reliable.
At higher helium abundances ($y \ga 2$), the physics of the fluid becomes increasingly challenging to model
and the synthetic spectra are more uncertain. 
These atmosphere models have been successfully applied in several studies of the SEDs of very cool white dwarf stars.
This includes two stars with near-UV spectra (BPM 4729 \citep{ks06} discussed in the introduction, and the WD companion to a pulsar \citep{durant12}),
studies of large samples of WDs from the SDSS or observed with {\it Spitzer} \citep{kilic08,kilic09a,kilic09b} as well as halo 
WD candidates \citep{hall08,kilic10b,kilic12}.

Model atmospheres and synthetic spectra were calculated for $\teff=4000$ to 6000$\,$K in steps of 100$\,$K,
$\log g \, ({\rm cm/s}^2) = 7$ to 9 in steps of 0.5, and for He/H number ratios of $y=-9$
(``pure H''), $-$2, $-$1, $-$0.3, 0, 0.3, 1, 2, 3, 4, 5, 6, and 20 (``pure He'').  Due to the 
difficulty of obtaining converged models for some combinations of parameters -- particularly for high gravity
combined with low-$\teff$ and high He/H ratios -- the final grid contains 969 of the 1365 models that
would completely cover the parameter space described above.
As we will see below, many combinations of parameters are irrelevant to our analysis and this partial
grid is amply adequate for our purposes.

Given the filter bandpasses we defined  for the STIS data (\S \ref{stis_obs}), the $UBVRI$ filters \citep{cohen03a}, the
2MASS $JHK_s$ filters \citep{cohen03b}, and the {\it Spitzer} IRAC 
bands\footnote{http://irsa.ipac.caltech.edu/data/SPITZER/docs/irac/calibrationfiles/spectralresponse/} 
we obtain synthetic photometric fluxes $F_i$ by integrating the synthetic spectra over model bandpasses
\begin{equation}
  F_i = \frac{\int_0^\infty f(\lambda) \lambda S_i(\lambda)\,d\lambda}{\int_0^\infty \lambda S_i(\lambda) \,d\lambda},
  \label{flux_int}
\end{equation}
where $f(\lambda)$ is the model flux density and $S_i(\lambda)$ the transmission curve of bandpass $i$.\footnote{The 2MASS
bandpasses are given as $\lambda S(\lambda)$.}  These synthetic photometric fluxes can then
be directly compared with the data.

\subsection{Fitting procedure}

The integrated model fluxes $F_i$ are  
fitted to the data by a least-squares method to obtain the stellar parameters. We proceed by finding the minimum of
the $\chi^2$ surface defined by
\begin{equation}
  \chi^2(\teff, \log g, y)= \frac{1}{N} \sum_{i=1}^N \Big[ \frac{f_i^{\rm obs} - (R(\teff,\log g)/D)^2 F_i(\teff,\log g,y) } 
                            {\sigma_i^{\rm obs}} \Big]^2
  \label{chi2}
\end{equation}
for a fixed value of $y$.
In Eq. (\ref{chi2}), $N$ is the number of bandpasses, $f_i^{\rm obs}$ the observed flux at the Earth in bandpass $i$, $\sigma_i^{\rm obs}$ its
uncertainty, and $D=1/\pi$ is the distance given by the parallax $\pi$. $R$ is the radius of the white dwarf 
obtained from the evolution sequences of \citet{fbb01}.  For a fixed value of $y$, Eq. (\ref{chi2}) defines
a surface in the $(\teff, \log g)$ plane that can be minimized to find the optimal parameters for each star.
A comparison of the fits and $\chi^2$ obtained for the different values of $y$ can then constrain the helium abundance
in the atmosphere. 

To fit a particular star, we first evaluate Eq. (\ref{chi2}) at each $(\teff, \log g)$ model grid point for a fixed value of $y$. Fluxes for
missing models are interpolated in $\teff$ with a cubic spline fit along a fixed gravity. Extrapolations are
sometimes necessary to fill the model grid, but fits falling in an extrapolated $(\teff,\log g)$ region are discarded. 
The filled $\chi^2(\teff,\log g)$ surface is fitted with a two-dimensional cubic spline for the purpose of interpolating
between grid points.\footnote{Our choice of spline interpolation was driven by the requirement that the interpolation be smooth. 
While a spline interpolation
along only five  points in gravity is not {\it a priori} reliable, we found that using a 5-point Lagrange
polynomial interpolation gives very nearly the same results. Piece-wise linear or quadratic interpolations
in $\log g$ bear a strong imprint of the $(\teff, \log g)$ grid and are not suitable.}  This function is then minimized using a downhill 
simplex algorithm \citep{num_recipes} to obtain the values of $(\teff, \log g)$ that best fit the data.  In all cases, 
the $\chi^2$ surface has a single, well-defined minimum (Figure \ref{chi2_surf}).

We determine the uncertainties on the fitted parameters by repeatedly fitting synthetic data generated by adding random variations 
around the measured values with a Gaussian distribution of a width given by the error bar. These synthetic data sets
are obtained by randomly sampling the distributioni of values for the parallax and each of the observed fluxes.
We found that 2000 such samplings give converged results. We thus obtain a distribution of
$(\teff, \log g)$ values for a fixed $y$ for each star. We take the average value and the dispersion
along $\teff$ and $\log g$ as the best-fit values and their uncertainties. An example of this procedure in shown in
Fig. \ref{chi2_surf}.  The $\chi^2(\teff,\log g)$ surface corresponding to the nominal observed value is shown by the contours. The minimum
is within a valley that corresponds closely with a line of constant bolometric luminosity (not shown).
The cluster of small dots results from the sampling of the observational error bars, and their average value
is indicated by the large dot and their standard deviation by the vertical and horizontal bars.  We find that the dispersion 
in gravity is primarily determined by the uncertainty in the parallax, while
that in $\teff$ reflects mostly the photometric error bars. In most cases, as in this example, the values of the gravity and
$\teff$ are only weakly correlated.

\section{Results}
   \label{results}

\subsection{Pure hydrogen models}

Based on previous analyses \citep{kilic09a,giam12}, we expect our target stars to have atmospheric compositions 
dominated by hydrogen. To cast our study in the context of the existing body of work, we first perform the analysis
with models of pure H composition ($y=-9$) which also provides a reference for the discussion of the He abundance
in these stars that will follow. 

We ascertain the validity of our fitting procedure by comparing with the recent work of
\citet{giam12} who have fitted the $BVRIJHK$ photometry of 169 nearby white dwarfs, including all 8 of our targets.
For this purpose, we fitted only the $BVRIJHK_s$ data (Tables \ref{tab_opt_fluxes} and \ref{tab_ir_fluxes}) with our pure H models.
This is consistent with their determination that all of our targets are better fitted with pure H atmospheres than
with pure He models.  Our $BVRI$ photometry is on the
same photometric system \citep{hb06} but we use 2MASS $JHK_s$ photometry while \citet{giam12}  
use CIT $JHK$ filters for six of our stars, and 2MASS photometry for the other two.  The adopted parallaxes are identical except
for WD 2054$-$050 which is slightly updated here but otherwise agrees with the value used by \citet{giam12} within the 
uncertainties.  We have verified that our pure
H models are nearly identical to those used in their analysis (P. Bergeron, priv. comm.). Thus differences
in the fitted $\teff$ and $\log g$ values can only arise from variations in the adopted fluxes in each band
(which are identical in many cases), 
differences in the adopted near-IR photometric systems, and the numerical implementation of the fitting procedure.

Figure \ref{comp_H} shows the excellent agreement between our results and those of
\citet{giam12}, well within their respective $1\sigma$ error bars in every case. Gravities 
agree to within 0.01$\,$dex
for 5 of the stars and differs by 0.04$\,$dex in the worst case (WD 0230$-$144).  The effective temperatures are also in 
excellent agreement
with an average absolute difference of 32$\,$K. There seems to be a systematic offset, however, as for all but one star
(WD 0747+073A) we find slightly higher effective temperatures.  These small
differences between our results for $\teff$ may be due to their coarser grid of models ($\Delta \teff=500\,$K, versus 100$\,$K here).
While our uncertainties in $\log g$ are essentially the same, those in $\teff$ are typically about half of those reported in \cite{giam12}. 
Since the uncertainty in our data (Tables \ref{tab_stis_fluxes}-\ref{tab_ir_fluxes}) is in nearly all cases the same as that reported in \citet{giam12},
this systematic difference in $\sigma_{\teff}$ must come from our different approaches to estimate the errors in the fitted parameters.

Having established that we reproduce recent published results very well, we now proceed to fit the entire
SEDs of our target stars, including the STIS data, $BVRI$ photometry, 2MASS $JHK_s$ photometry and the 
fluxes in all four {\it Spitzer} IRAC bandpasses with pure H models. 
This set of 14 bandpasses spans wavelengths from 0.24 to 9.3$\,\mu$m, which includes 90--96\% of
the bolometric flux of these cool stars. As $\teff$ and $\log g$ are the only two free parameters in the models (we only
consider pure H models for now), this data set strongly constrains the fits and constitutes a stringent test of the models.

The $\teff$ and gravities obtained from fitting the SEDs with pure H models ($y=-9$) are given in Table \ref{tab_fits} 
and in Figure \ref{comp}. Compared to the values we obtained by fitting only the optical and 
near-IR photometry only, the $\teff$ have increased for five stars and decreased for the other three,
but generally within the uncertainties. A similar pattern is found for the change in $\log g$.  Although two stars
(WD 1444$-$174 and WD 2054$-$050)  stand out for their much larger change in the values of both $\teff$ and $\log g$, there is no
pattern in terms of gravity, $\teff$ or He/H ratio (see below) in the effect of adding the near-UV and the IRAC fluxes to 
the fitting procedure. On the other hand, the near-UV and mid-IR fluxes very effectively constrain the range of models that
can fit the data, resulting in formal uncertainties that are 2--3 times smaller than when fitting only the optical and near-IR fluxes
(Figure \ref{comp}). The typical uncertainty is 10--20$\,$K, 
which shows the advantage of using as broad a photometric baseline as possible to accurately determine the
$\teff$ of the coolest DA and DC white dwarfs. This is only the formal internal uncertainty and it does not include 
errors due to the limited fidelity of the models or in the photometry that are systematic in nature. The 
true uncertainty on $\teff$ is likely larger than the values 
quoted in Table \ref{tab_fits}.  On the other hand, the uncertainty in the gravity is unchanged as it depends primarily on the
accuracy of the parallaxes.

Except for WD 0230$-$144, for which the $\chi^2$ is almost as
good as for the fits with the optical and near-IR fluxes only, fits of the complete SEDs have larger 
values of $\chi^2$.
This is not surprising as it is much harder to get models to fit a full SED. It also shows that while
our description of the physics of these very cool atmospheres is fairly realistic (see below), it needs to be further
refined to fully match the precision of the data.  Nonetheless, some very good fits are
obtained with this pure H composition for WD 0230$-$144 and WD 0357+081 (Figure \ref{best_sp1}).
In the next section we discuss the overall best fit for each star.

\subsection{Mixed helium and hydrogen models}

Several mechanisms of great import occur in the atmospheres of cool WDs such as inter-species diffusion, convective mixing, 
and the accretion of interstellar or
planetary material.  These processes cause temporal variations in the atmospheric composition that, when combined with 
the changes in the chemistry and opacity as $\teff$ decreases lead to distinctive changes in WD spectra, collectively known
as spectral evolution.
Because of decreasing thermal excitation, absorption lines of neutral He disappear in WD spectra 
for $\teff \lesssim 10000\,$K \citep{bwf91, tremblay10} and the presence of He in the atmosphere must be inferred indirectly.
In cool DA stars for instance, only H lines are seen but the presence of neutral He atoms can broaden the Balmer line profiles \citep{wehrse77,lw83,bwf91}.
In the coolest stars with H-rich atmospheres ($\teff \lesssim 4800\,$K), the Balmer lines also become invisible for lack of thermal excitation
and their spectra are featureless \citep{brl97}, the defining characteristic of the DC spectral class.  In the absence of spectral 
lines or with only a very weak H$\alpha$ line, the spectroscopic method \citep{bwf91,bsl92} can no longer be applied and
the composition of these very weak DA stars and very cool DC stars must be inferred primarily from the shape
of their SEDs.  Continuum absorption processes involving H$_2$ and He each have characteristic wavelength
dependences that can, in principle, reveal the composition of the atmosphere. Two opacity sources are of particular interest 
to this study.  In the near-UV, the absorption is dominated by the red wing of the Ly$\alpha$ line which is pressure-broadened  
primarily by collisions with H$_2$ and H, but 
also with He \citep{ks06}.  The latter is the weaker broadening mechanism but it can become important for large He/H
ratios. The other absorption mechanism is the collision-induced absorption (CIA) by H$_2$ molecules \citep{af13}  which absorbs
mainly in the near-IR with a particularly strong band in the $K$ bandpass. The very broad roto-translational band of H$_2$ CIA peaks
around 5--10$\,\mu$m and affects the mid-IR flux. In cool DA and DC white dwarfs, the H$_2$ CIA occurs whether the collisional partner is
another H$_2$ molecule, a H atom or a He atom. In He-enriched atmospheres, the lower H$_2$ abundance tends to reduce the CIA,
an effect that is over compensated by the higher photospheric density in such models, resulting in very strong H$_2$ CIA for $y \sim 2-5$
\citep{bsw95,bl02}.  Our data set spans these two critical spectral
regions and offers a valuable opportunity to constrain the He/H ratio in very cool DA and in H-rich DC.

In the previous section, we fitted the data with pure H models by minimizing $\chi^2(\teff, \log g)$.
By varying the He/H ratio, it may be possible to improve the quality of the fit 
and determine the He abundance in these atmospheres \citep{kilic09a}. We have repeated the fitting procedure
for every value of He/H ratio in our model grid and studied the behavior of the minimum $\chi^2$ as a function
of $y$ (Figure \ref{heh}). The general trend is that the $\chi^2$ is relatively constant for small He/H ratios,
which reflects the fact that the model spectra are not sensitive to small amounts of He.  For $y \wig>0$,
the $\chi^2$ rises rapidly to large values.  In several stars, there is a minimum around $y \sim -0.3$. Our model
grid covers values of $y$ from $-9$ to 20 but only the lower range is shown in Figure \ref{heh}. For large
mixing ratios of $y\sim 2-5$, the CIA of H$_2$ due to collisions with He becomes extremely strong \citep{bl02}, 
causing a severe flux depression in the near-IR that  is not seen in our target stars. This drives the $\chi^2$ to very large
values and leads to distorted $\chi^2(\teff,\log g)$ surfaces and anomalous fitted stellar parameters. This range of mixing ratios
is clearly irrelevant to our sample.  When H becomes a trace element in the atmosphere ($y \wig> 5$), the H$_2$ CIA is considerably
reduced and the $\chi^2$ values fall steadily to reach a low value for pure He models.

Given our method of sampling the observational uncertainties, we not only obtain distributions for $\teff$ and the gravity,
but also of $\chi^2$, as each set of simulated data gives its own minimum value of $\chi^2$. This allows
a determination of  the significance of the minima seen in Figure \ref{heh} and to estimate the uncertainty on the He/H ratio.
The standard deviation of the $\chi^2$ distribution for each value of $y$ is shown as an error bar in Figure
\ref{heh}. We consider that differences in $\chi^2$ that are less than $1\sigma$ are not significant and that the
corresponding fits are not distinguishable given the observational uncertainties. Specifically, for two He/H ratios
$y_1$ and $y_2$ with dispersions $\sigma_1$ and $\sigma_2$, if
\begin{equation}
  |y_2-y_1| \le \sqrt{\sigma_1^2 + \sigma_2^2}
  \label{heh_crit}
\end{equation}
we consider the two values to be equally probable.  For example, in the case of WD 0230$-$144 in Figure \ref{heh},
it is not until $y=-0.3$ that $\chi^2(y)$ rises far enough above the value of $\chi^2(y=-9)$ (pure H model) to be distinguishable
based on this criterion. Thus, we find an upper limit of $y\le-0.3$ to the helium abundance
in the atmosphere of this star. Table \ref{tab_fits} gives the He/H ratio that corresponds to the lowest $\chi^2$, the corresponding
values of $\teff$ and $\log g$ (technically the best fitting values) as well as the range of composition allowed by the uncertainties 
in the data, rounded off to the nearest 0.1 dex.

Most studies of the SEDs of very cool white dwarfs of spectral types DA and DC have assumed either
pure H or pure He composition. If we restrict our analysis to the optical and near-IR fluxes, as in many previous studies, we find 
that one star in our sample (WD 0230$-$144) is fitted equally well by a pure H and a pure He model.  The pure He solution 
is significantly different, however, with $\teff$ lower by 247$\,$K and a gravity that is 0.18$\,$dex lower. 
For all the other stars, fits with pure H ($y=-9$) models
always have a lower $\chi^2$ but for four stars (WD 0357+081, WD 0657+320, WD 1444$-$174 and WD 1820+609),
the pure He models fits are equally good given the above criterion (Equation \ref{heh_crit}). This is consistent with the photometric
fits shown in \citet{giam12} for these stars, in particular for WD 1820+609 where their pure H and pure He models
appear to fit the $BVRIJHK$ fluxes equally well.  In this case, as in many others, the
presence of a weak H$\alpha$ line indicates a star with a H-rich composition, thus lifting the ambiguity in the solution 
obtained by fitting the photometric fluxes.  We note that \citet{giam12} found all eight stars in our sample to have a pure H composition.   

On the other hand,
our fits of the full complement of photometric measurements clearly show that a pure H composition provides
a far better fit than a pure He composition for the eight stars in our sample. Thus we conclude that none have atmospheres of a pure
or dominant He composition, including the three stars of spectral type DC.  Not surprisingly, we further find that for fits based only on the 
optical and near-IR photometry ($BVRIJHK$), the sensitivity of the $\chi^2$ on the He/H ratio is much weaker compared to that
shown in Figure \ref{heh} and generally weaker constraints on the composition (higher upper limits) are found.
This emphasizes the value of a wide wavelength coverage that includes the near-UV an the mid-IR 
for the analysis of the SED and the determination of the atmospheric composition in particular. A broad wavelength coverage is 
especially valuable for the DC stars that do not show any Balmer lines or for stars without H$\alpha$ spectroscopy.

In general, fits involving more free parameters tend to match data more closely and thus to reduce the $\chi^2$. It would not
be surprising then that our fitting procedure which, in addition to the usual $\teff$ and $\log g$, considers $\log {\rm He/H}$ as
a free parameter would result in all of our stars containing some amount of He. The implication would be
that 1) they do indeed have atmospheres of mixed composition, 2) that adding some He compensates for either some inadequacy in the
constitutive physics of the models or systematic errors in the photometry (such as the absolute calibration across the SED),
or 3) a combination of the above. Notwithstanding these general considerations, 
including the He/H ratio as a free parameter is not merely a numerical exercise because the model fits
are strongly constrained by the wide wavelength coverage of our data. The addition of He to a hydrogen atmosphere affects the modeled SED
in specific ways that do not necessarily correct the limitations in the models and the data. It is thus very different from 
performing least-squares fits of data with polynomials, for example, where higher order polynomials 
invariably produce fits with lower $\chi^2$. Our eight stars show a range of behavior in $\chi^2$ as a function of composition that builds confidence 
in the validity of the He/H ratios we derive.  For WD 0230$-$144, WD 0357+081 
and WD 1444$-$174, $\chi^2(y)$ is constant or {\it increases} steadily as $y$ increases 
(Figure \ref{heh}).  In these three cases, adding He does not improve the fit to the data,
the best fits are obtained for a pure H composition and we are able to set upper limits to the He abundance.
Three stars ( WD 0657+320, WD 1820+609 and WD 2054$-$050) show a very shallow minimum for $-1 \lesssim y \lesssim 0$ that is not 
significant and  we set upper limits on the He/H ratio for these stars.
Finally, both stars in the pair WD 0747+073AB  show clear minima in $\chi^2(y)$. The most extreme case is WD 0747+073B
whose $\chi^2$ is reduced by nearly a factor of 2 when going from a pure H composition to $y=0$.

\subsection{Best fits of the spectral energy distributions}

So far we have determined the best values (with uncertainties) of $\teff$ and of the gravity as well as obtained 
upper limits, and in one case, a limited range of values for the He/H ratio (Table \ref{tab_fits}).  
The model SEDs giving the lowest $\chi^2$
(boldfaced entries in Table \ref{tab_fits}) are shown for each star in Figure \ref{best_sp1}.
The residuals
of each fit are shown scaled to the uncertainty of each photometric measurement. A comparison of those same models
with the STIS spectra is shown in Figure \ref{stis_fits}. We discuss each star individually before discussing
trends and drawing conclusions. The relatively recent studies of \citet{kilic09a} and \citet{giam12}  have analyzed our target 
stars with the photometric method and models that include the Ly$\alpha$ line profiles of \citet{ks06}. In the following
discussion we compare our results for individual stars with theirs as appropriate.

\subsubsection{\it WD 0230$-$144}  
At $\teff=5528\,$K, this DA is at the upper limit of $\teff$ in our sample and is remarkably well-fitted by a pure
H model ($\chi^2=1.6$). Eleven of the 14 bandpasses are fitted within their $1\sigma$ error bars. 
In particular, the STIS fluxes are fitted very well. Our $\teff$ is slightly higher than that determined by \citet{giam12} (5477$\,$K).
While a pure H composition provides the best fit of this star, we find that concentrations of up to $y=-0.3$ are allowed by the
photometry.  At this upper limit of the He abundance the effective temperature would be 5612$\,$K. 

\subsubsection{\it WD 0357$+$081}  
This star is slightly hotter than WD 0230$-$144, with $\teff=5565\,$K (\citet{giam12} find a nearly
identical 5578$\,$K). The fit is also excellent, with 7 bands fitted within $1\sigma$ and the worst mismatch is for the STIS3 flux at
$1.7\sigma$.  Considered together, the STIS fluxes are reproduced very well.
The SED of this DA is best reproduced by a pure H model but we find that He abundances of up to
$y = -0.4$ are allowed by the uncertainties, which would drive $\teff$ up to 5633$\,$K.

\subsubsection{\it WD 0657$+$320} 
Formally, the best fit  of the SED of this DA is obtained for a He abundance of $y=-1$ ($\chi^2=2.8$)
but a pure H model gives a fit that is essentially just as good ($\chi^2=3.1$, see Figure \ref{heh}).  At $\teff=4999\,$K,
(\citet{giam12} find $\teff=4888\,$K), this star is cooler than the previous two and the overall fit is not quite as good. Seven of the
fourteen fluxes are fitted within $1\sigma$ and the outliers are the $R$ flux ($-2.7\sigma$) and the IRAC [3.6] flux ($2.8\sigma$).
The STIS fluxes are reproduced by the model to better than $1.3\sigma$.  We find an upper limit on the He abundance of $y\le 0$.

\subsubsection{\it WD 0747$+$073A}  This is the faintest and coolest star in our sample. 
As discussed in \S \ref{stis_obs}, we strongly suspect that this star was not properly acquired in the slit of the STIS
instrument. Additional evidence comes from our fits of the SED including the upper limits in the STIS fluxes. The best fitting
model matches the STIS1 upper limit but overestimates the STIS2 and STIS3 upper limits by factors of 3 and 2.6,
respectively. This models ($\teff=4293\,$K, $\log g=7.80$, $y=-0.3$) predicts STIS2 ands STIS3 fluxes that are about half of those
of the best fitting model for WD 2054$-$050 (see below), which should have been easily detected in view of the 11$\sigma$ and 6.3$\sigma$
signals for the latter. Thus, unless the near-UV flux of WD 0747+073A is strongly suppressed by an unidentified absorber, our analysis
is consistent with our empirical arguments for having observed blank sky.

Excluding the upper limits for the three STIS bandpasses from the analysis of WD 0747$+$073A, we find that the 
fit of the $BVRIJHK_s$ and IRAC fluxes with pure H models is distinctly
worse than for the three previous stars.  Allowing for a mixed composition clearly improves the quality of the fit as it reduces the
$\chi^2$ from 5.4 to 2.8. The
best match to the data is found for $y=-0.3$, giving $\teff=4354\,$K.  The well-defined minimum in the $\chi^2$ (Figure \ref{heh})
leads to a determination of the helium abundance of $-1 \wig< y \wig< -0.1$. 
Figure \ref{wd0747a} shows the
best fitting models with pure H composition and with $y=-0.3$. The pure H model (open circles) underestimates the $R$ flux
by $3.3\sigma$ and overestimates the IRAC [3.6], [4.5], and [5.8] fluxes significantly (up to $3.5\sigma$).
The addition of helium to the model improves the optical region of the fit significantly, as well as the first three IRAC fluxes.
It also increases the H$_2$ CIA absorption in the near-IR, improving the match with the $J$ flux while becoming too strong in the $K$ band, however.
\citet{giam12} find no trace of H$\alpha$ absorption
in the spectrum of this star, which is consistent with both our proposed solution with a mixed composition and their best fit with a low-$\teff$ model
of pure H composition.  \citet{kilic09a} also find a pure H composition for this star on the basis of the $BVRIJHK_s$ and 
IRAC fluxes. As both \citet{giam12} and \citet{kilic09a} include the
Ly$\alpha$ profiles of \citet{ks06}, the principal difference between the models of these two studies and those used here
is our updated H$_2$-He CIA opacity \citep{abel12} which may very well explain why we obtain a better fit with a mixed He/H composition. 

\subsubsection{\it WD 0747$+$073B} 
This star is brighter than its companion WD 0747$+$073A in all bands 
and published analyses consistently find that it is hotter by a few hundred degrees. 
Like its companion, this star was undetected in the STIS observations.  If we include the upper limits on
the STIS fluxes in the analysis, we obtain a rather poor fit with $\chi^2=8.7$ for a model with a mixed He/H atmosphere with
$y=0$.  As for WD 0747+073A, the STIS upper limits are systematically overestimated by 2.0$\sigma$, 6.4$\sigma$ and 4.3$\sigma$, respectively,
and the near-UV fluxes predicted by the best fitting model SED should have been readily detectable.  WD 0747+073B present the same situation as
its dimmer companion and we discard the upper limits on the STIS fluxes for the same reasons.
Fitting the eleven remaining bandpasses, a pure H model reproduces the SED reasonably well, with $\chi^2=3.5$. The $B$ and $R$ 
fluxes are poorly matched however, with the model deviating from the data by $2.5\sigma$ and $-3.3\sigma$, respectively. Increasing the
He abundance improves the fit in nearly every band (Figure \ref{wd0747b}) and the lowest $\chi^2$ is obtained for $y=0$,
with a $\teff$ of 4723$\,$K and a lower gravity of $\log g=7.95$. The latter values are close to those
reported by \citet{giam12}. The relatively high He abundance causes the H$_2$ CIA to be too strong in the $K_s$ band however.  
The minimum in $\chi^2(y)$ (Figure \ref{heh}) gives an allowed range of $-1.5 \le y\le 0.2$
using the criterion given by Equation \ref{heh_crit}. If this criterion is relaxed slightly,  even a pure H model gives
an acceptable fit within the uncertainties (Table \ref{tab_fits} and Figure \ref{wd0747b}).  The presence of a significant amount of
helium in this star's atmosphere is consistent with the extremely weak H$\alpha$ detection shown in \citet{giam12}.  Fitting the $BVRIJHK$ and
IRAC fluxes, \citet{kilic09a} found $\teff=4700\,$K, $\log g=7.97$ and $y=-0.1$, which is in excellent agreement with our result.

\subsubsection{\it WD 1444$-$174}
This is the highest gravity star in our sample ($\log g=8.49$) and its SED is very well reproduced by a pure H model except for the $B$ filter
where the model overestimates the observed flux by a spectacular $6.4\sigma$. The other model fluxes agree with
the data to within $1\sigma$ in nearly all bands (Figure \ref{best_sp2}), with particularly good matches of the neighboring STIS1 and $V$
fluxes.  The $\chi^2$ for this star increases 
steadily with the He fraction (Figure \ref{heh}). A pure H composition is thus favored with $\teff=5205\,$K.  
The large mismatch of the $B$ flux is rather puzzling. The adopted $B$ flux (Table \ref{tab_opt_fluxes}) combines two measurements
and agrees within $1\sigma$ with the value used by \citet{giam12}. If we exclude the $B$ band from the fit, pure H models give essentially
the same result. On the other hand, if we exclude the STIS data and fit all the other bands (including $B$), $\teff$ decreases substantially
to 5043$\,$K with pure H models, and the best fit is obtained for $y=0$ at $\teff=4937\,$K and a reduced gravity of $\log g=8.35$. The latter values are
in good agreement with those of \citet{giam12} for a pure H model. From these experiments, we conclude that the STIS data drive the fit to high $\teff$.
Since there is no reason to suspect that the STIS and $B$ fluxes for this star are erroneous, there remains the possibility that the Ly$\alpha$
profile is too weak in the very far wing probed by the $B$ bandpass. While this is the highest gravity star in our sample, 
it does not stand out in terms of the physical conditions at the photosphere.  The photospheric density of the model is rather 
typical for these stars and comparable to that of the pure H model that gives a good fit of WD 1820$+$609 (see below). 
It is difficult to understand how a model that works quite well (see \S\ref{disc}) would do poorly in this particular instance.
Nonetheless, the overall fit of the SED of WD 1444$-$174 is quite reasonable (Figure \ref{best_sp2}).

Ignoring the difficulty fitting the $B$ flux, we find an upper limit of $y=-0.7$ for this star, with a corresponding $\teff=5242\,$K.
Fitting the $BVRIJHK$ and IRAC fluxes, \citet{kilic09a} report a much lower $\teff=4820\,$K and that the SED is best 
matched with a mixed composition of $y=-0.15$, which is well above our $1\sigma$ upper limit.
The spectrum of WD 1444$-$174 shown in \citet{giam12} shows no trace of H$\alpha$ absorption, while their model for the comparable star WD 0552$-$041 
($\teff=5182\,$K, $\log g=8.37$, pure H composition) shows a very weak line. The absence of the line in WD 1444$-$174 could be caused by 
the presence of He which would broaden the line into invisibility. This suggests that the atmosphere of WD 1444$-$174 has a mixed composition
closer to our upper limit of He/H $\sim 0.2$ than to pure hydrogen. On the other hand, the lack of H$\alpha$ is easily accommodated if
$\teff$ is lowered as reported in \citet{giam12} and \citet{kilic09a}.

\subsubsection{\it WD 1820$+$609}
A pure H model provides a reasonable fit to the SED of this star, with $\chi^2=4.0$, but it misses the STIS1 flux ($-2.8\sigma$), the
STIS3 flux (3.1$\sigma$), and the $R$ flux ($-2.9\sigma$). While the slope of the model SED is slightly steeper than that of the data
in the near-UV,
the overall behavior of the STIS data is reproduced very well (Fig. \ref{stis_fits}).  Adding He to the model barely improves the fit. The formally best
fit is obtained for $y=-1$ (Figure \ref{best_sp2}, for which the agreement with the data is slightly better in every band.  An upper 
limit of $y=-0.1$ can be placed on this star. Interestingly, its $\teff$ and gravity are barely affected by
including the near-UV and the mid-IR data in the analysis or by adding He. Our $\teff$ and gravity for this star ($\teff=4921\,$K, $\log g=7.96$)
are essentially identical to those of \citet{giam12} and \citet{kilic09a}.

\subsubsection{\it WD 2054$-$050}
This is the second coolest and the lowest gravity star in our sample with $\teff=4491\,$K and $\log g=7.84$ -- similar to
those of WD 0747+073A (Table \ref{tab_fits}).  As we found for WD 1444$-$174, including the near-UV and mid-IR fluxes in the analysis of WD 2054$-$050
significantly increases $\teff$ (+160$\,$K) and the gravity (+0.19 dex) while assuming a pure H composition.  This pure H fit
is of modest quality, matching 6 of the 14 bands within $1\sigma$, but missing the $I$ and 
overestimating all four IRAC fluxes, especially the [3.6] band. However, the IRAC photometry of this star is likely contaminated 
by a nearby bright star \citep{kilic09a}.  The STIS fluxes are all reproduced with $1\sigma$. A marginally better fit is
obtained for $y=-0.3$ (Figure \ref{best_sp2}) which brings the model into much better agreement with the data in the $R$, $I$ and IRAC [3.6] and 
[4.5] bandpasses, but at the cost of spoiling the match with the STIS1 and  STIS2 fluxes by steepening the slope of the model in the near-UV.  
The H$_2$ CIA absorption becomes too strong at this helium abundance and the $K_s$ flux is clearly too low.
While our upper limit on the He abundance is fairly high at $y=0.1$, a pure H or H-dominated composition (He/H $<< 1$)
is more likely (with $\teff=4517\,$K and $\log g=7.84$), given the uncertain IRAC fluxes. At this low $\teff$, the absence of the H$\alpha$ 
line in this DC star is consistent with both a pure H composition \citep{giam12} and our upper limit of $y=0$.

To summarize, we find that we can obtain good to excellent fits of our target stars over their full SEDs, from 0.24 to 9.3$\,\mu$m. The extension
of the fits to the near-UV with the STIS observations strongly constrain the models. 
We are able to provide realistic upper limits on the He/H ratio in the atmospheres, which are typically $N({\rm He})/N({\rm H}) < 0.1 - 1$.  

The pair of stars WD 0747+074AB stands out from the rest of the sample.  Unexpectedly, neither was detected with STIS and our models
fit the corresponding flux upper limits poorly. We have presented theoretical and empirical evidence that the non-detection 
is most likely due to a failure to acquire those two targets in the STIS aperture. 
An alternate explanation could be that both stars in this pair have peculiar atmospheres with another
near-UV absorber in addition to the red wing of Ly$\alpha$. Since component A is a DC and the component B a DA with extremely  
weak H$\alpha$ \citep{giam12}, there is no direct evidence for the presence of any element other than H and He in their atmospheres.
The fact that we obtain good fits of their SEDs when we exclude the STIS upper limits suggests that neither star is particularly unusual.
While the lack of STIS data for these two stars is unfortunate, we were able to perform our analysis with the 11 remaining bandpasses
and found good evidence that WD 0747+073A has a mixed He/H atmosphere and is thus the most interesting star in our sample.

\section{Discussion}
   \label{disc}

Our primary goal was to apply a stringent test to the opacity models for the red wing of the Ly$\alpha$ line of
\citet{ks06} in very cool hydrogen-rich WDs. {\it HST} STIS spectroscopy probes the steeply rising 
near-UV spectrum that is a combination of
the Ly$\alpha$ profile and the Wien tail of the SED at these very low $\teff$. This steep slope provides a sensitive test of
the theory. By choosing stars for which the SED is already very well characterized from the optical to the mid-IR and  with known distances,
we strongly constrained the range of $\teff$ and gravity that could accommodate our new data.  For five of the six stars with STIS data,
our fits of the entire SED are good to excellent matches of the blue and near-UV spectral range.  However, the model has difficulty 
simultaneously matching STIS fluxes of WD 1444$-$174 and its $B$ flux.  Overall, this constitute very strong evidence 
that this absorption mechanism is indeed the ``missing blue opacity'' of earlier models \citep{brl97} and that
the model for the broadening -- primarily involving collisions with H$_2$ -- is on a firm quantitative basis
even though there is evidence that a refined theoretical model could be even more successful.
Earlier analyses of the SEDs of very cool WDs that use the \citet{ks06} Ly$\alpha$ profile 
\citep{ks06,kilic09a,kilic09b,giam12} further support this conclusion. The validation of this opacity model
is important as it directly affects the determination of the ages of globular clusters from their WD sequence
\citep{kowalski07}, for example.

The overall quality of the fits of the full SEDs is not uniform and ranges from excellent to satisfactory.
While we cannot reliably identify trends from a sample of eight stars, we note that the fits
tend to become worse at lower $\teff$ and higher gravity.  These are the stars with 
higher photospheric densities and more extreme physical conditions. This suggests that 
models of the microscopic physics of dense matter should be further refined.  In particular, we found that to
match the near-UV fluxes of some stars the presence of helium was necessary but the corresponding increase
in the strength of the H$_2$ CIA in the near-IR is larger than what is allowed by the data. This tension
between the Ly$\alpha$ line wing and the CIA absorption indicates that one or perhaps both
mechanisms are responsible for the mismatch.  For example, the static broadening model of \citet{ks06} could be
improved by considering dynamic collisions \citep{ak09} or more accurate potential and dipole surfaces for
the H-H$_2$ system during a collision.  Another possibility is the corresponding ``Ly$\alpha$'' absorption from
a H$_2$ molecule between its electronic ground state and its first electronic excited state, which is not accounted for
in the \citet{ks06} model. This process
could become important at low $\teff$ where hydrogen is mostly recombined into H$_2$ molecules and the atomic 
Ly$\alpha$ line would disappear.  The calculation of the H$_2$-He CIA has recently been completely
revised \citep{af13} but a comparable update of H$_2$-H$_2$ CIA in the regime of WD atmospheres is still in progress 
(M. Abel, pers. comm.). In the relatively dense atmospheres that have He/H $\ga 1$, CIA from triple collisions 
(e.g. H$_2$-H$_2$-He, H$_2$-He-He) may play a role but is completely unexplored in the relevant conditions.
These limitations in the current models must introduce systematic biases in the determination of
$\teff$, $\log g$ and the composition that are nearly impossible to quantify, unfortunately.

\begin{figure}
   \plotone{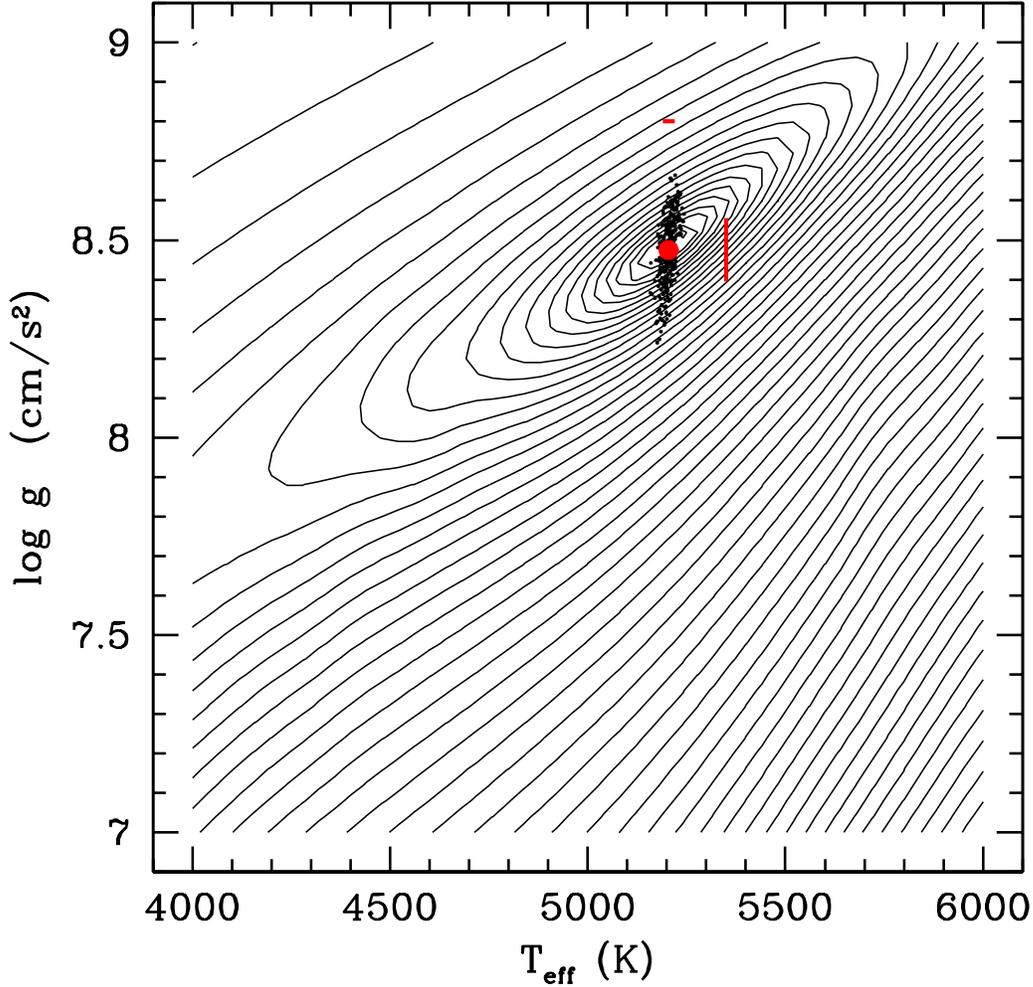}
   \caption{Illustration of the fitting procedure. The star is WD 1444$-$174 and the composition is
            fixed at $\log N({\rm He})/N({\rm H})=-9$ 
            (``pure H'').  The surface of $\chi^2(\teff,\log g)$ between the models and the nominal parallax
           and observed fluxes is shown by contours that are evenly spaced by an arbitrary amount. 
           The well-defined minimum of the surface gives a value of ($\teff$, $\log g$). 
           Monte Carlo sampling of the uncertainties in the data generates similar but different $\chi^2$ surfaces
           whose minima are shown by the cluster of small dots.  The centroid and the
           dispersion of this cluster of solutions give the adopted values of $\teff$ and $\log g$ 
           (large red dot) and their dispersions (red bars, offset for clarity): $\teff=5205 \pm 14\,$K and $\log g=8.49 \pm 0.08$. 
           See the text for details.  In this example,
           the major and minor axes of the cluster of solutions are nearly aligned with the
           ordinate and the abscissa, so the dispersions in $\teff$ and $\log g$ are uncorrelated. 
           [{\it See the electronic edition of the Journal for a color version of this figure.}]}
    \label{chi2_surf}
\end{figure}

\begin{figure}
   \epsscale{1.00}
   \plotone{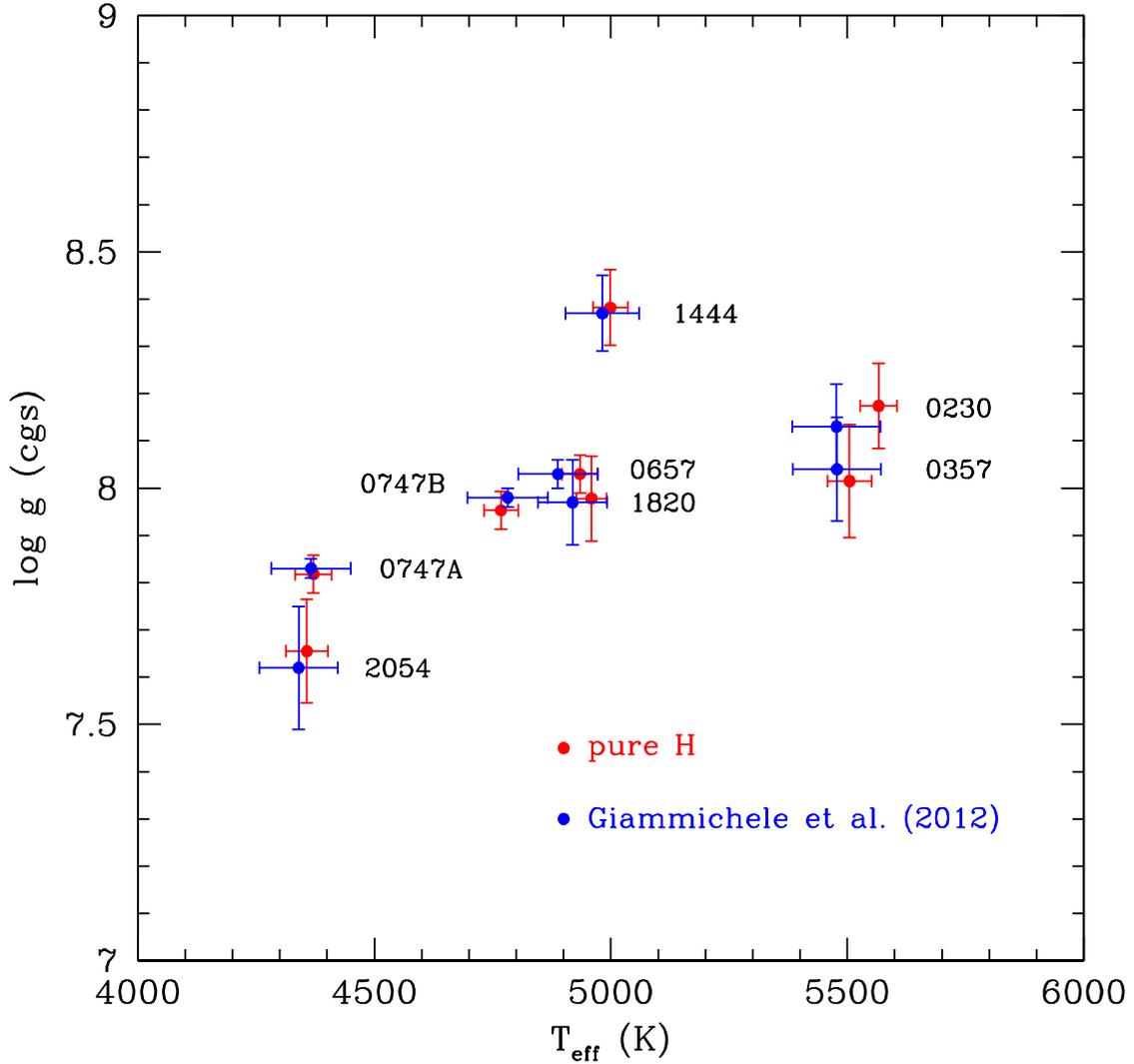}
   \caption{Effective temperature and gravities for our 8 target stars (red symbols) assuming a pure hydrogen composition.
            In this figure, only the $BVRIJHK_s$ fluxes were fitted for comparison 
            with the similar work of Giammichele et al. (2012) (blue symbols).  
            Matching pairs of red and blue symbols for each star are labeled with the first four digits of the WD designation. 
           [{\it See the electronic edition of the Journal for a color version of this figure.}]}
    \label{comp_H}
\end{figure}

\begin{figure}
   \epsscale{1.00}
   \plotone{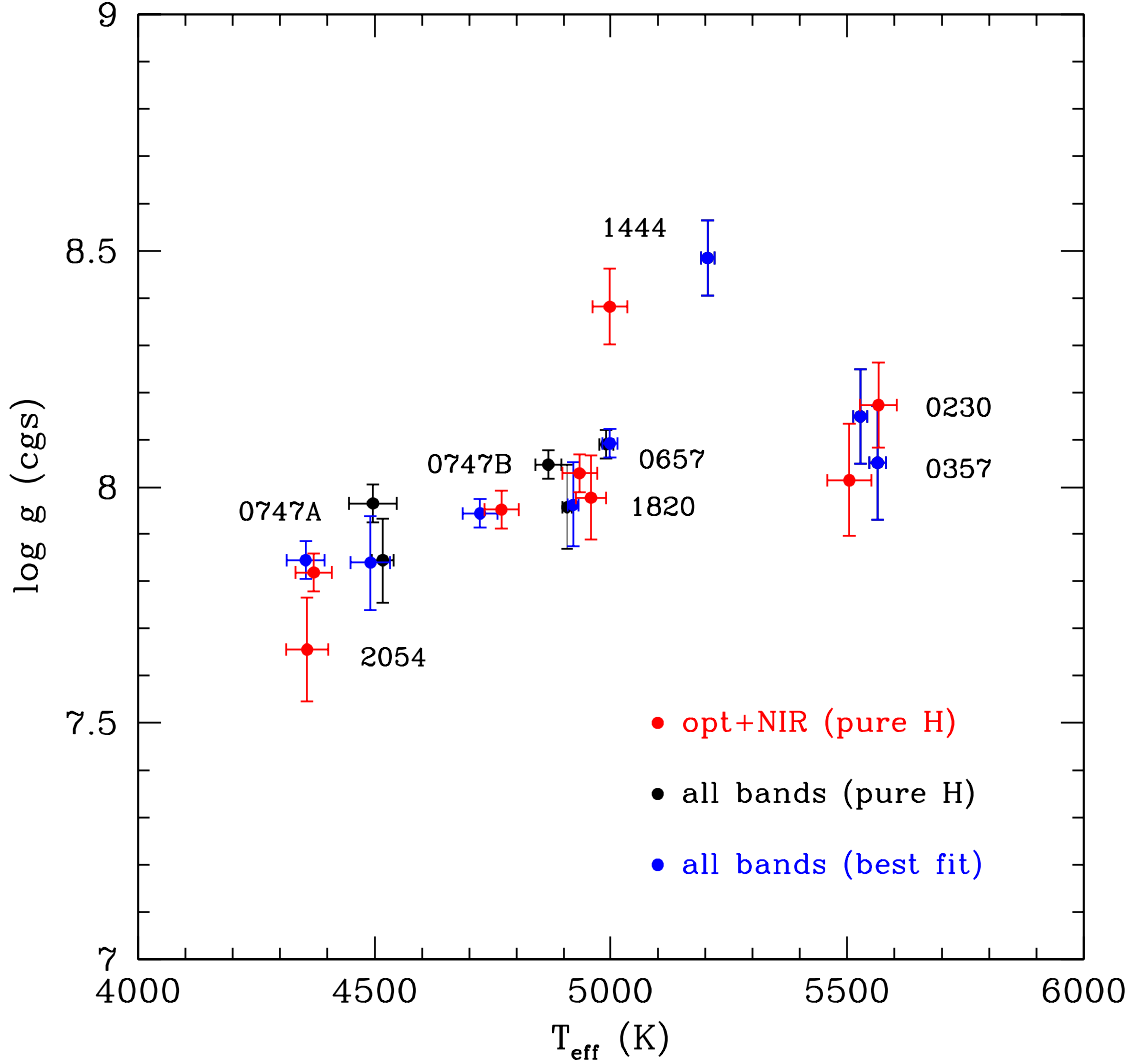}
   \caption{Effective temperature and gravities for our 8 target stars.  Fits of the optical and near-infrared
            photometry only, assuming a pure H composition are shown in red and correspond to those shown 
            in Figure \ref{comp_H}.
            Values obtained by fitting the full SEDs (STIS, $UBVRIJHK_s$ and IRAC fluxes), with pure H composition
            are shown in black.  The best fitting parameters  obtained when allowing the He/H ratio
            to vary are shown in blue.  Triplets of symbols (red, black, blue) are
            labeled with the first four digits of the WD designation. In some cases, pure H is the
            best fitting composition and the black and blue symbols overlap.  See Table \ref{tab_fits} for
            the numerical values. For the pair of stars WD 0747+073AB, the upper limits on the STIS fluxes
            were excluded from the fitting procedure (see text).
           [{\it See the electronic edition of the Journal for a color version of this figure.}]}
    \label{comp}
\end{figure}

Most determinations of the parameters of cool white dwarfs with the photometric method
employed here are based on optical and near-IR fluxes and achieve typical uncertainties in $\teff$ of $\sim 100\,$K \citep{kilic10a,giam12}. 
For example, \citet{giam12} find an average uncertainty of $\pm 84\,$K for the stars in our sample.
By extending the measured SEDs of our targets in the near-UV with the {\it HST} STIS data and including the mid-IR IRAC
fluxes, we very effectively constrain the model fits with
a considerably reduced uncertainty averaging $\pm 24\,$K (Table \ref{tab_fits}). Excluding the STIS fluxes typically
doubles the $\teff$ uncertainty. As we discussed earlier, the
gravity of these stars is primarily determined by the parallax and our determinations are close to
those of previous studies that used the same parallaxes (Table \ref{tab_stis_fluxes}).
Compared to \citet{giam12}, the average gravity of our sample is higher by 0.04 dex but agrees perfectly with that of \citet{kilic09a}.
On the other hand,  our $\log g$ uncertainties are identical for each star as they are dominated by the parallax uncertainties.
Much more precise determinations of the gravity and of the bolometric luminosity as well will come from the 
Gaia mission, which will be able to measure the parallaxes of $\sim 10^4$ WDs to 1\% accuracy \citep{carrasco14}. For relatively
bright, nearby WDs such as those in our sample, a much higher precision of $\sigma_\pi \la 60\,\mu$as is anticipated, or about 
0.1\% \footnote{http://www.cosmos.esa.int/web/gaia/science-performance}, compared to the current 1.3 - 7.1\% (Table \ref{tab_stis_fluxes}). 
However, experimentation with our fits reveal that
the uncertainty on the gravity becomes dominated by the photometric uncertainties when $\sigma_\pi$ is reduced by a factor of $\sim 10$
from the current values (i.e. to $\sim 0.5$\%) leading to $\Delta \log g \sim 0.01 - 0.02$. More accurate gravities
will then require more precise photometry. For the few stars where the $\teff$ and $\log g$ distributions are correlated (unlike
WD 1444$-$174 shown in Figure \ref{chi2_surf}) more precise parallaxes also lead to more precise $\teff$. 
The ground-based photometry used in this paper is characterized by uncertainties on the order of 0.03 magnitudes,  a common standard for most
photometric observations.  Fortunately, future prospects for
routinely improving ground-based photometry are bright.  The Large Synoptic Sky Survey (LSST) is designed to cover half the sky from the
southern hemisphere in the $ugrizy$ bands\footnote{http://www.lsst.org/lsst/scibook}.  Its design goals are 1\% absolute
and 0.5\% relative photometry.  Of particular interest here is the $u$ band
which would cover much of Johnson $U$ band that had to be abandoned in this analysis as unreliable
and the availability of the $y$ flux would help better characterize the SEDs in the far-red.  The combination of the Gaia and LSST high
precision data for nearby, very cool WDs would reduce the uncertainty in $\teff$ by a factor of 2--3 and to $<0.01$ dex in $\log g$.
Such data will highlight the limitations of the models and challenge the theoretical description of the microphysics of WD atmospheres.

The atmospheric composition of cool WDs is of central importance in the context of their spectral evolution. White dwarfs that
have a hydrogen surface layer at high $\teff \sim 20000\,$K can become He-dominated at lower $\teff$ because of convective mixing with the
underlying He layer that is at least $10^2$ times more massive \citep{dm79,fw97}. The depth of the surface hydrogen convection zone increases rapidly at
$\teff \la 12000\,$K and again for $\teff \sim 6000\,$K to bottom out around 5000$\,$K (see for example Figure 1 of \citet{tb08}).
Evolution models predict that at its deepest extent, the H convective zone amounts to $\sim 10^{-6}$ of the mass of a typical
WD of $M_\star =0.6\,M_\odot$ \citep{fbb01}.  Thus, DA white dwarfs with a hydrogen layer thinner than $10^{-6}M_\star$ are expected
to become He dominated with He/H $>> 10^2$ before they cool below $\sim 5000\,$K. According to this convective mixing
mechanism, stars with thicker layers never mix and
retain a H-rich atmosphere throughout their low-$\teff$ cooling, turning into H-rich DC stars below $4800\,$K.
There is strong empirical evidence supporting this convective mixing process, with an estimated $\sim 15$\% of DA having
hydrogen layers thin enough to mix convectively with helium \citep{tb08,giam12}.

Most very cool WDs have atmospheres dominated by either H (spectral types DA and DC) or He (spectral types DZ, DQ, DQpec
and DC).  \citet{ks06} found that very few, if any, very cool WDs ($\teff \la 6000\,$K) of types DC have helium-rich atmospheres.  
If we exclude the ultracool WDs that are very poorly understood, few studies of very cool DA and DC white dwarfs with the photometric 
method have considered models of mixed composition. 
In their analysis of $BVRIJHK$ and IRAC photometry of 43 DA and DC WDs (including all of our targets), \citet{kilic09a}  found 11 stars
that were better fitted with mixed atmospheres with ${\rm He/H} \sim 1$  ($y=-0.7$ to 0.3) but did not provide uncertainties 
on the fitted parameters.  This study is the most comparable to ours
as it uses a broad coverage of the SED (but no near-UV data) and similar model spectra including
the Ly$\alpha$ profiles of \citet{ks06}. In a subsequent study based on a large sample of new very cool WDs discovered in the 
SDSS, \citet{kilic10a} fitted $grizJHK$
photometry and H$\alpha$ spectroscopy with models that did not include the Ly$\alpha$ absorption and without the benefit of parallaxes. 
They considered mixed He/H composition only for stars that showed large near-IR flux deficits, which typically are caused by strong H$_2$--He CIA.
This approach tends to restrict the determination of mixed He/H composition to ultracool WDs, and indeed, they find only one among
their subset of DA and DC stars with $y\sim -1$ and 17 others with very high He abundances of $y=1$ -- 6. 

The determination of the He abundance in a very cool DA or DC star is rather challenging as
their SED is rather insensitive to the presence of small amounts of He. For instance,
the difference in flux between a pure H model and one with $y=-1$ is at most 0.4\% at $\teff=6000\,$K (with $\log g=8$) and
grows to 1.1\% at 5000$\,$K. Only for $\teff \la 4500\,$K do the calculated SEDs depart from each other noticeably, by $\sim 7$\% 
in the near-IR. Given that typical photometric uncertainties in the optical and near-IR are $\sim 3$--5\%, it
becomes clear that the photometric method of analysis is insensitive to He abundance of He/H $\la 0.1$, except for
the coolest stars. Furthermore, there is evidence that small systematic shifts between the calibrations of the optical and near-IR
photometry of cool white dwarfs remain (P. Bergeron, priv. comm.) that could affect determinations of the He/H ratio.
Better constraints will require significantly more precise and accurate photometry \footnote{For very cool DA 
with known parallaxes however, it may be possible to complement the 
photometric method with an analysis of the weak H$\alpha$ to constrain the He abundance, but not for DC stars, of course.}
as well as improvements in the models.
Our method of analysis allows for the first time the determination of upper limits on the abundance of
He in stars with H-rich atmospheres and provide error bars for the one star where the atmosphere is of mixed composition.
In all but one star, we found that a pure H composition provided a fit that is statistically equivalent to the
best fit composition. For these stars the  upper limits on the helium abundance ranges from $y=-0.7$ to 0.2.  These He/H ratios
of order unity are consistent with the overall results of \citet{kilic09a}.  The most interesting star in our sample is the DC WD 0747+073A
for which we find that a pure H model is not compatible with the data and we can constrain the helium abundance to $-1 \le y \le -0.1$.
At $\teff=4354\,$K, this is the coolest star in our sample and we would
expect that the photometric method would be sensitive to the He/H ratio. Thus, our detailed analysis is consistent with earlier
studies that found that some very cool WDs of type DC can have mixed He/H atmospheric composition with He/H $\sim 1$. 
This is unlike the ultracool WDs that have ratios estimated to be $10^2$ -- $10^6$ \citep{kilic10a} or the DZ stars
whose atmospheres are dominated by He with traces of metals and He/H $> 10^3$ \citep{dufour07}.
Moderate He/H ratios are unexpected in the context of the convective mixing scenario where the H layer eventually mixes
and become dominated by helium (He/H $> 10^2$) or the H layer is so thick that mixing never occurs and a pure H surface composition persists.
Perhaps WD 0747+073A is a rare example of a star with an initially He-rich atmosphere that has been moderately enriched in H by accretion from
the interstellar medium \citep{tremblay14}.

The WD cooling ages of our stars based on the best fit parameters (Table \ref{tab_fits}) and the evolution sequences of \citet{fbb01}
(with a H surface layer of $10^{-4}M_\star$) range from 3.4 to 8.1$\,$Gyr and
are generally very close to those of \citet{giam12} and well within the error bars.
We find that the formal uncertainty on the age is dominated by the uncertainty in the gravity for all of our stars.
Two stars merit further discussion. For WD 1444$-$174, the effects of our larger gravity and larger $\teff$ largely compensate each 
other for a net decrease in the cooling age of 0.5$\,$Gyr.  Only for WD 2054$-$050 is there a noticeable change in age, 
which increases from 4.3 to 5.9$\,$Gyr which is just within the $1\sigma$ uncertainties.  
As much as can be ascertained from this small sample and considering their relatively large uncertainties, 
the ages of very cool hydrogen-rich disk white dwarfs in the range of $\teff \sim 4300$ -- 6000$\,$K are not biased by 
using $\teff$ and gravities derived from optical and near-IR photometry only.

\acknowledgments

We thank the referee,  Pierre Bergeron, for his constructive comments and sharing his insights and Gilles Fontaine for providing
tables of WD evolution sequences. Support for this publication was 
provided by NASA through program number HST-GO-12188.02-A from the Space Telescope Science Institute, which is operated by the 
Association of Universities for Research in Astronomy, under NASA contract NAS5-26555.
This work is based in part on observations made with the Spitzer Space Telescope, which is operated by 
the Jet Propulsion Laboratory, California Institute of Technology under a contract with NASA.

\bibliographystyle{apj}
\bibliography{references}

\clearpage


\clearpage

\clearpage

\clearpage

\clearpage

\begin{figure}
   \epsscale{1.0}
   \plotone{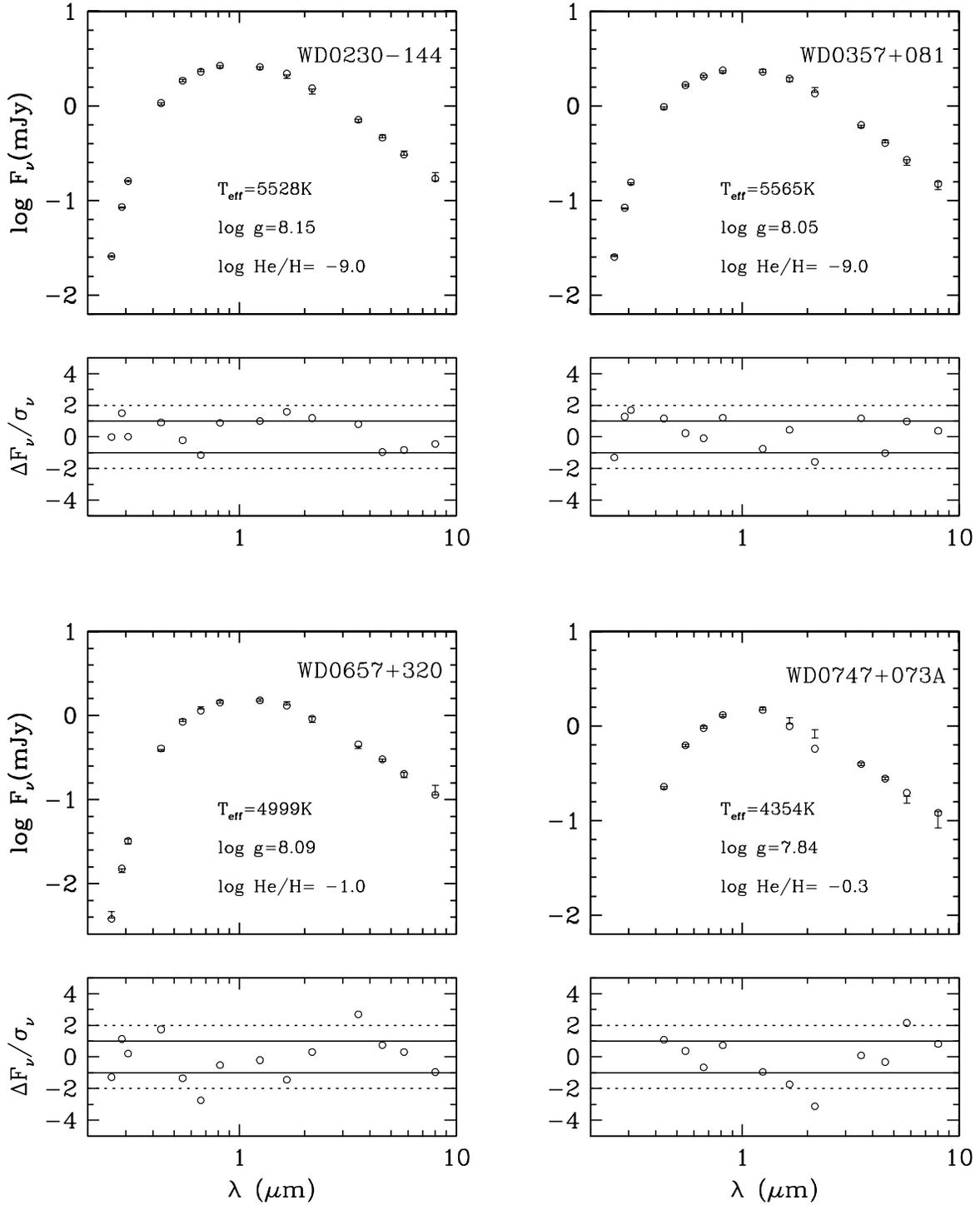}
   \caption{Comparison of the best fitting models with the data.
            For each star, the upper panel shows the model parameters (Table \ref{tab_fits}), the photometry with
            $\pm 1\sigma$ error bars and the model photometry with open circles.  The lower panel shows the
            residuals of the fit (model $-$ data), normalized to the uncertainty in each band.
            For the pair of stars WD 0747+073AB, the upper limits on the STIS fluxes
            were excluded from the fitting procedure (see text).}
    \label{best_sp1}
\end{figure}
\clearpage

\begin{figure}
   \figurenum{\ref{best_sp1}}
   \epsscale{1.0}
   \plotone{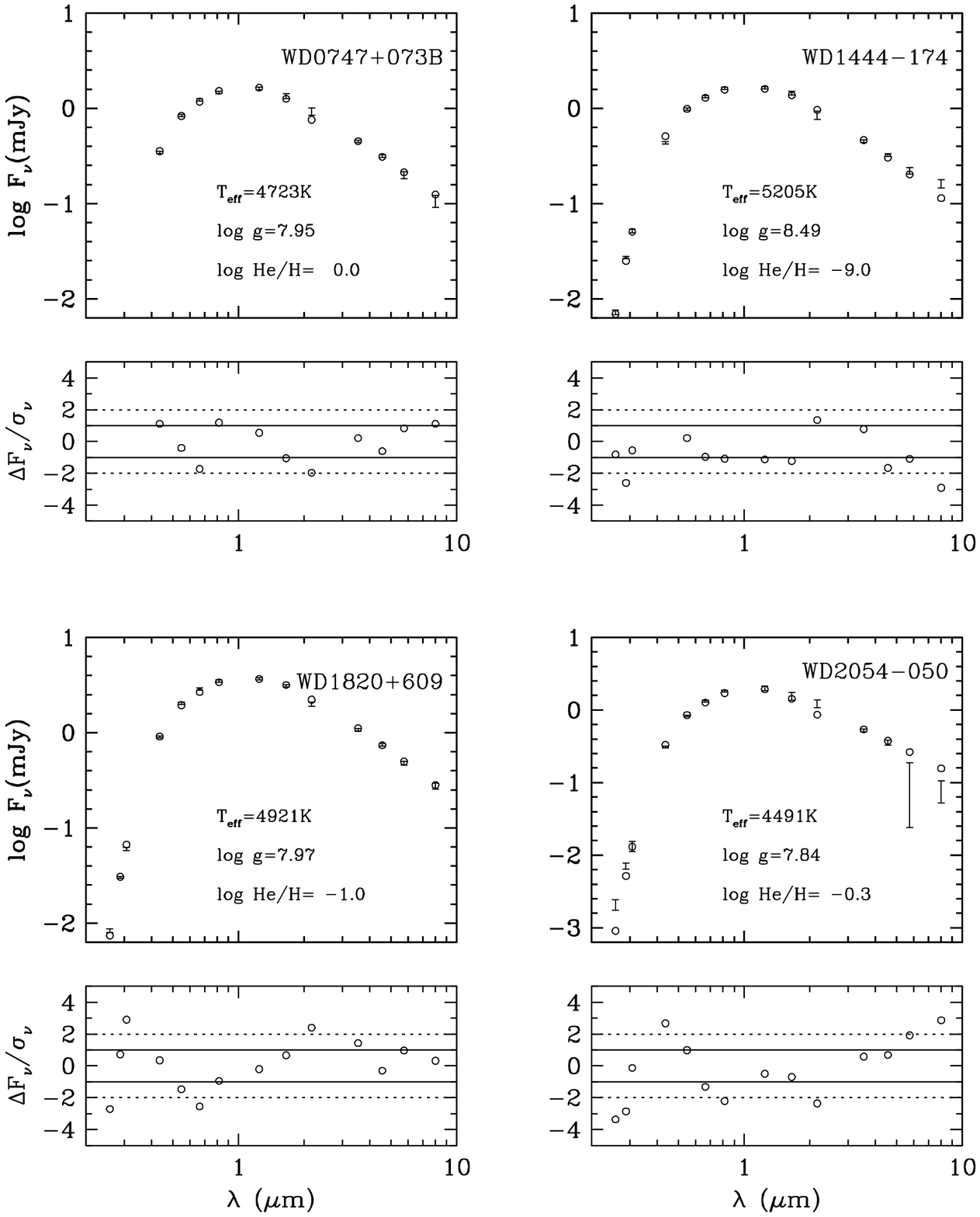}
   \caption{(Continued)}
    \label{best_sp2}
\end{figure}
\clearpage

\begin{figure}
   \epsscale{0.95}
   \plotone{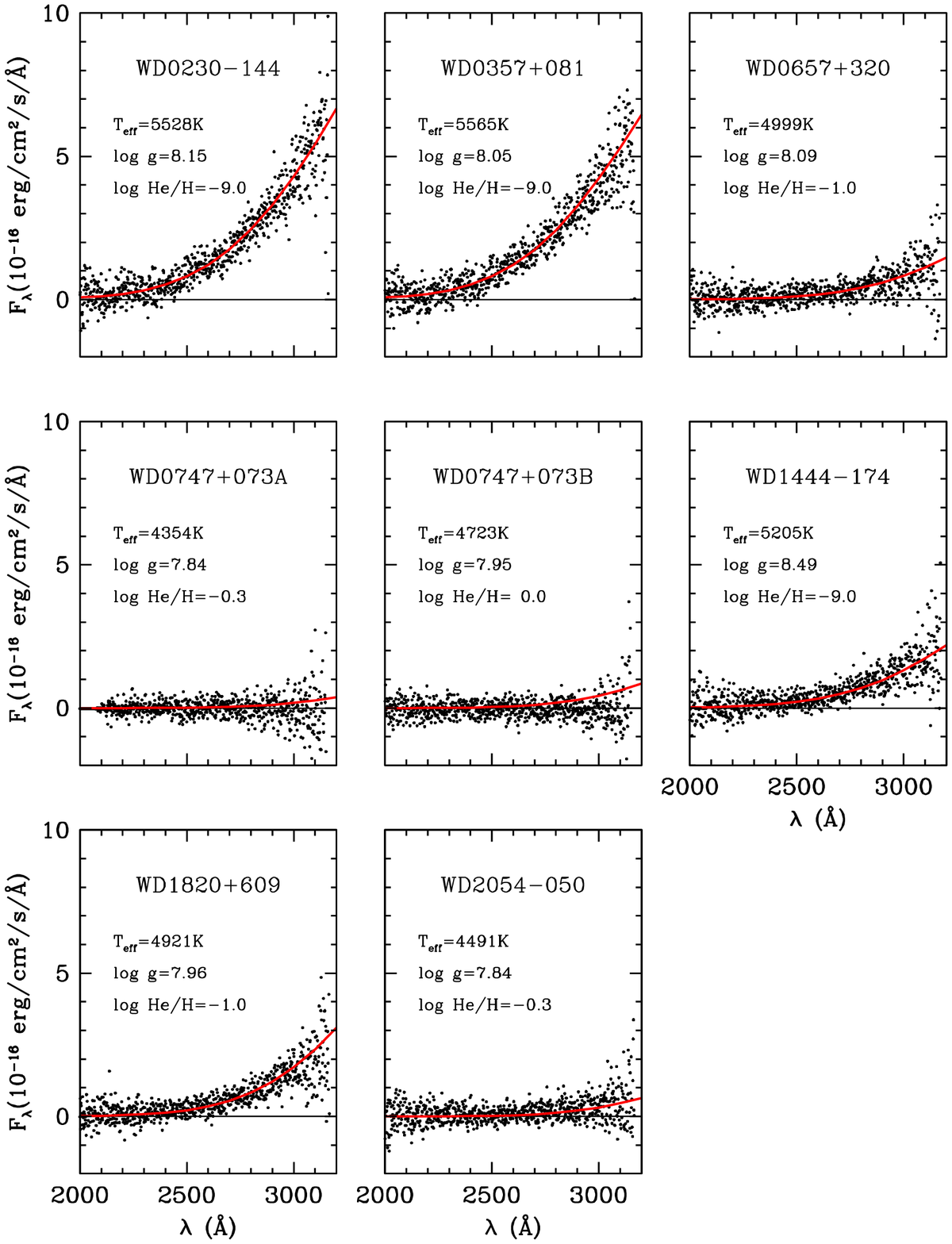}
   \caption{Comparison of the best fitting models (red curves)  with the STIS spectra (black dots)
            shown without error bars for clarity (see Fig. \ref{stis_sp}). 
            The model spectra are the same as those in Fig. \ref{best_sp1}.
            [{\it See the electronic edition of the Journal for a color version of this figure.}]}
    \label{stis_fits}
\end{figure}
\clearpage

\begin{figure}
   \epsscale{1.1}
   \plotone{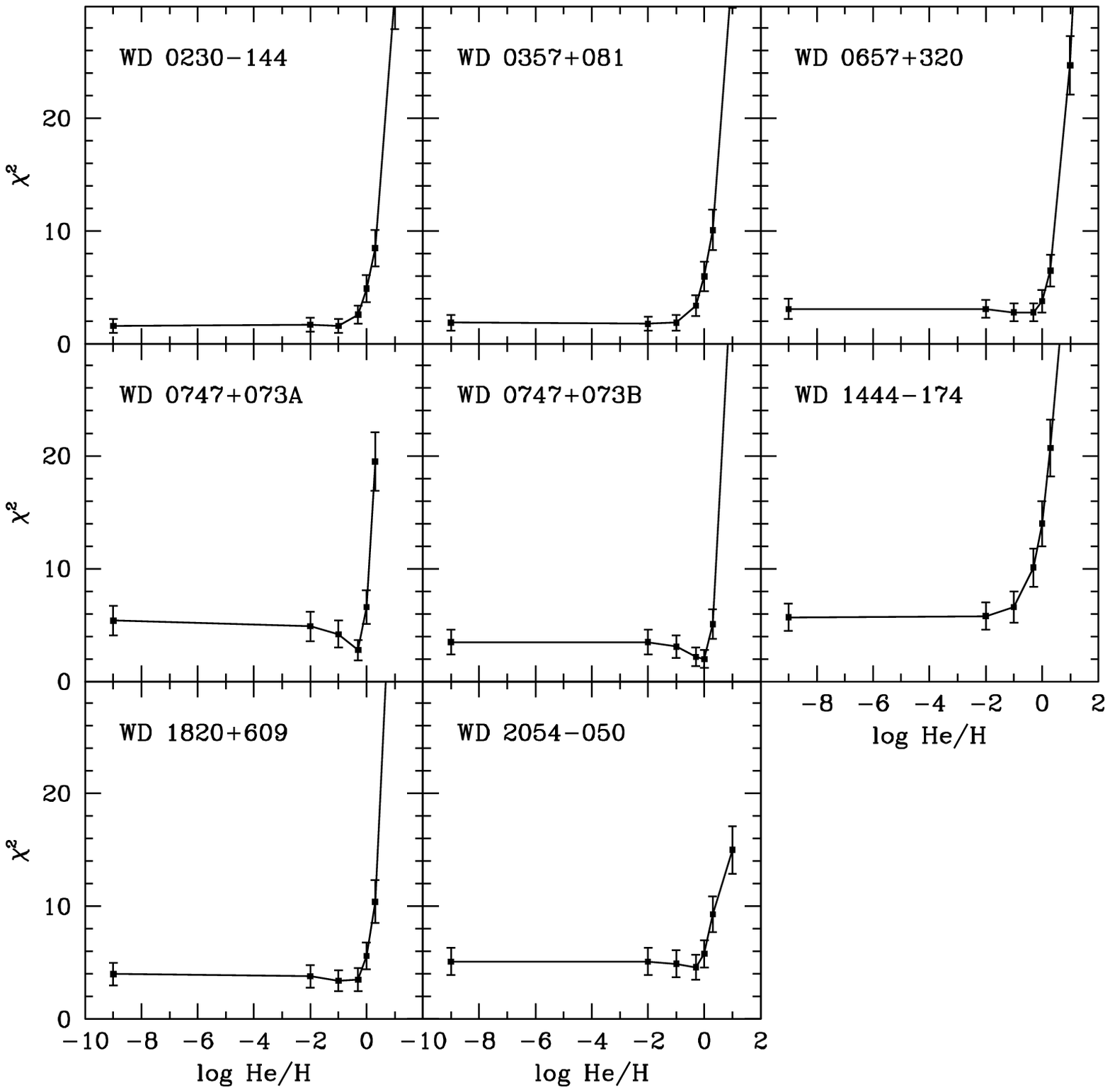}
   \caption{Behavior of the $\chi^2$ of the best fit model as a function of the He/H ratio.
            The lowest $\chi^2$ for each star indicates the best fitting He/H ratio.
            For the pair of stars WD 0747+073AB, the upper limits on the STIS fluxes
            were excluded from the fitting procedure (see text).  See Table \ref{tab_fits}.}
    \label{heh}
\end{figure}
\clearpage

\begin{figure}
   \plotone{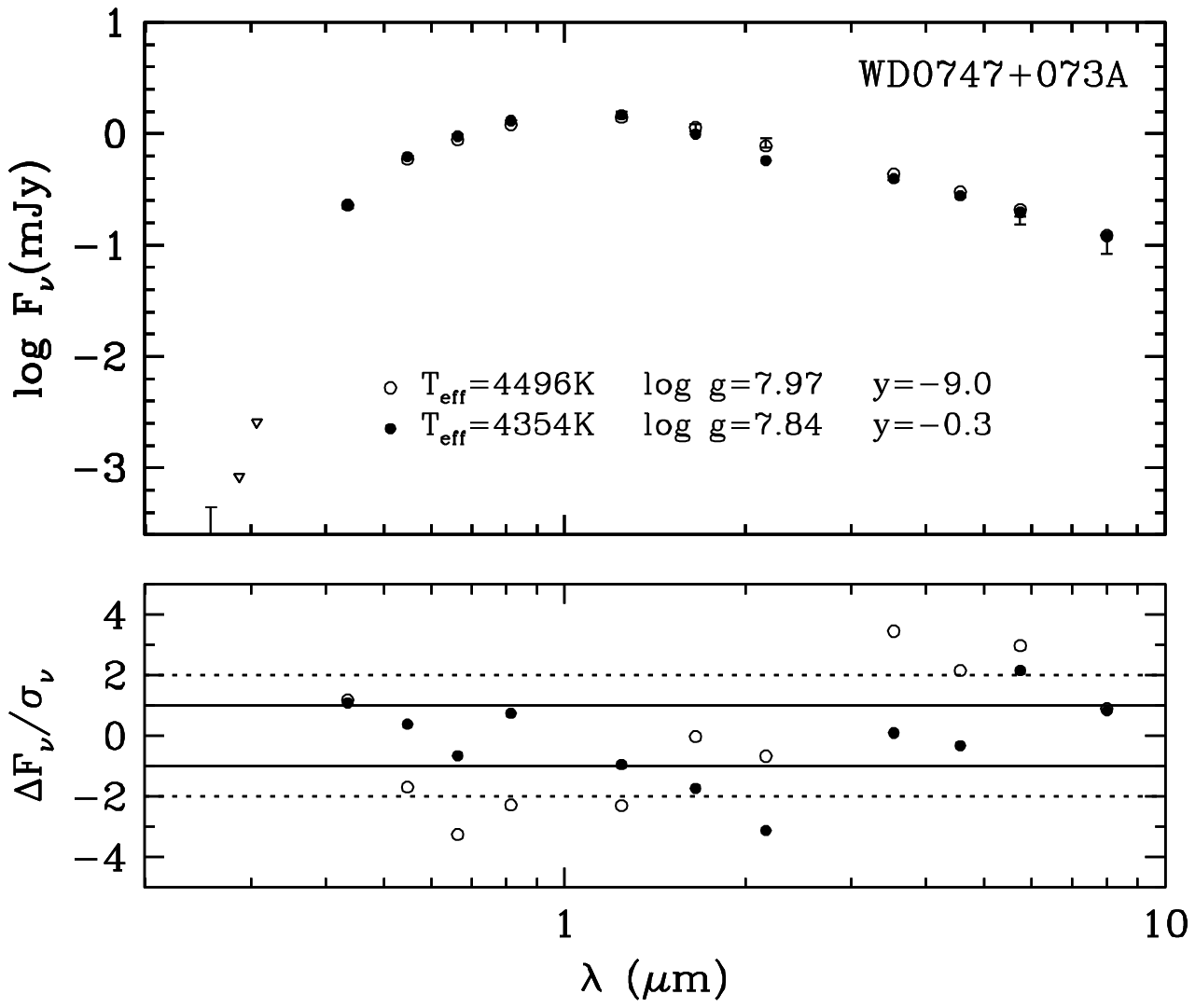}
   \caption{Two fits of the SED of WD 0747+073A, one with pure H composition ($y=\log {\rm He/H}=-9$,
            open circles) and the fit with the lowest $\chi^2$ ($y=-0.3$, solid circles). The lower
            panel shows the residuals, each normalized to the observational uncertainty. While the
            upper panel shows the upper limits on the STIS fluxes reported in Table \ref{tab_stis_fluxes}
            (inverted triangles), they were not included in the fits shown here (see text).}
    \label{wd0747a}
\end{figure}
\clearpage

\begin{figure}
   \plotone{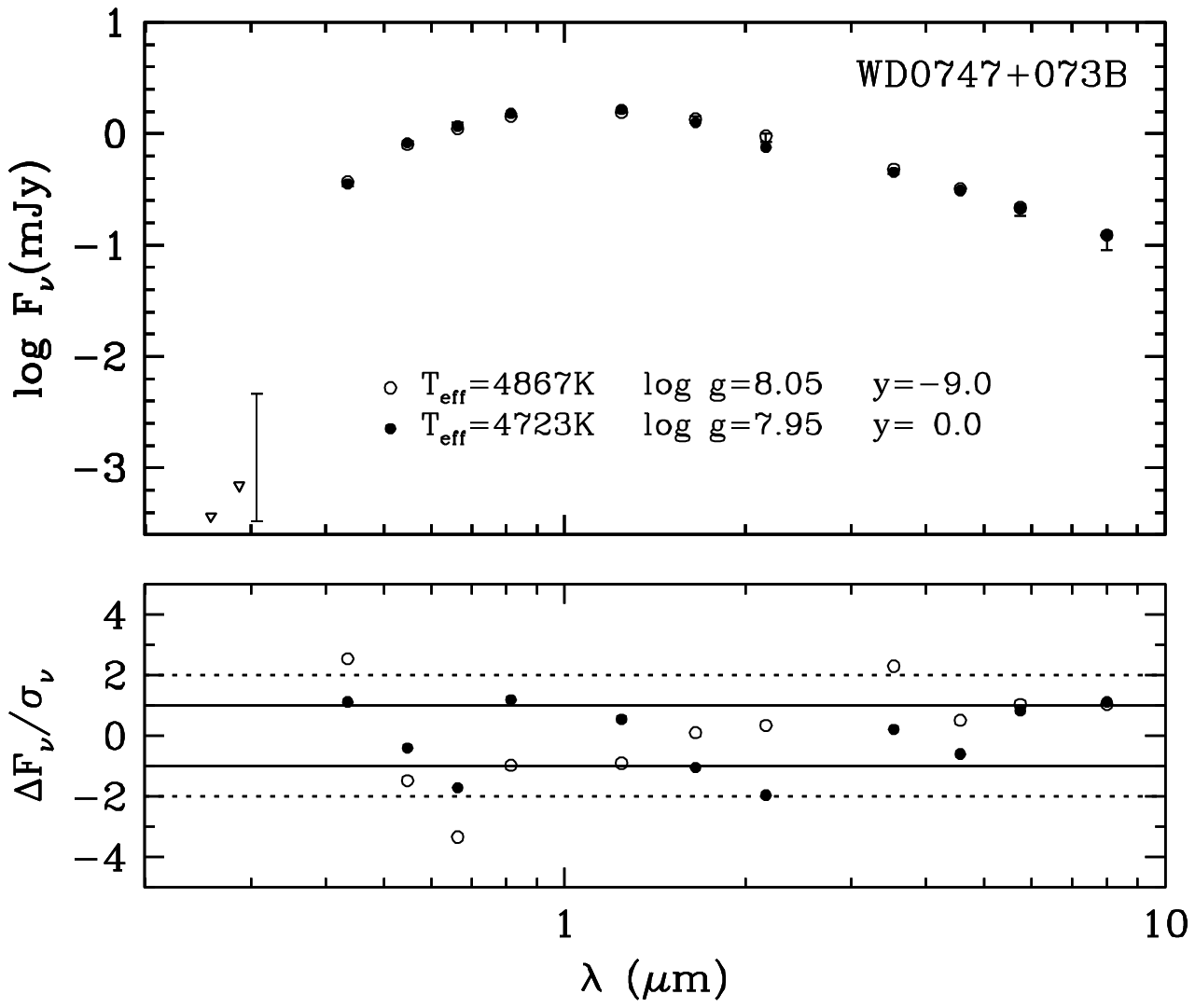}
   \caption{Two fits of the SED of WD 0747+073B, one with pure H composition ($y=\log {\rm He/H}=-9$,
            open circles) and the fit with the lowest $\chi^2$ ($y=0$, solid circles). The lower
            panel shows the residuals, each normalized to the observational uncertainty. While the
            upper panel shows the upper limits on the STIS fluxes reported in Table \ref{tab_stis_fluxes}
            (inverted triangles), they were not included in the fits shown here (see text).}
    \label{wd0747b}
\end{figure}
\clearpage

\begin{deluxetable}{lcccccccc}
\tabletypesize{\scriptsize}
\tablewidth{0pt}
\tablecaption{Adopted STIS photometric fluxes\tablenotemark{a} }
\tablehead{
\colhead{WD} & \colhead{GJ} & \colhead{Spectral}  &  \colhead{$\pi$} & \colhead{$\sigma_\pi$} & \colhead{STIS1}  &  \colhead{STIS2}  &  \colhead{STIS3} &  \colhead{Ref.\tablenotemark{b}}  \\
\colhead{}  & \colhead{}  & \colhead{type}  &  \colhead{(mas)} &   \colhead{(mas)}}
\startdata

0230$-$144  &  3162   &  DA  &  64.00  &   3.90  &    117.0 $\pm$  2.11  &    306.0 $\pm$  3.70  &    512.0 $\pm$  9.59    &  1     \\
0357+081    &  3259   &  DA  &  56.10  &   3.70  &    118.0 $\pm$  2.21  &    301.0 $\pm$  3.82  &    482.0 $\pm$  9.80    &  1     \\
0657+320    &  3420   &  DA  &  53.50  &   0.90  &     \phantom{1}19.5 $\pm$  1.70  &     \phantom{1}52.0 $\pm$  2.63  &    101.0 $\pm$  7.28    &  1     \\
0747+073A\tablenotemark{c}   &  1102B  &  DC  &  54.70  &   0.70  &      $<1.70$          &      $<3.03$          &       $<8.18$  & 1       \\
0747+073B\tablenotemark{c}   &  1102A  &  DA  &  54.70  &   0.70  &      $<1.68$          &      $<2.53$          &      \phantom{11}7.9 $\pm$  6.81  & 1    \\
1444$-$174  &  3866   &  DC  &  69.00  &   4.00  &     \phantom{1}32.9 $\pm$  1.73  &     \phantom{1}98.4 $\pm$  2.81  &    165.0 $\pm$  7.51    &  1     \\
1820+609    &  4054   &  DA  &  78.20  &   4.10  &     \phantom{1}38.2 $\pm$  1.66  &    110.0 $\pm$  2.73  &    192.0 $\pm$  7.37    &  1     \\
2054$-$050  &  812B   &  DC  &  59.53  &   3.10  &      \phantom{11}9.5 $\pm$  1.60  &     \phantom{1}25.7 $\pm$  2.42  &     \phantom{1}42.4 $\pm$  6.76    &  1, 2  \\   

\\
\enddata
\tablenotetext{a}{Fluxes are in $10^{-18}\,$erg/s/cm$^2/$\AA \ at Earth. See text for the definition of the STIS$x$ bandpasses.
                  Upper limits are 1$\sigma$ values.}
\tablenotetext{b}{Sources of Parallaxes: 1. \citet{YPC}, 2. \citet{hipparcos_new}.}
\tablenotetext{c}{The upper limits in the STIS fluxes appear to be from observations of blank sky.}
\label{tab_stis_fluxes}
\end{deluxetable}

\begin{deluxetable}{lcccccc}
\tabletypesize{\scriptsize}
\tablewidth{0pt}
\tablecaption{Adopted optical photometric fluxes\tablenotemark{a} }
\tablehead{
\colhead{WD} &  \colhead{GJ}  &  \colhead{$B$}  &  \colhead{$V$} &  \colhead{$R$}  & \colhead{$I$}  & \colhead{Ref.\tablenotemark{b}}\\
}
\startdata

0230$-$144  &  3162   &  1680.0 $\pm$ 46.40  &   1880.0 $\pm$ 52.00  &   1670.0 $\pm$ 46.00  &   1190.0 $\pm$ 32.90    & 1, 2   \\
0357+081    &  3259   &  1520.0 $\pm$ 41.90  &   1670.0 $\pm$ 46.10  &   1450.0 $\pm$ 40.10  &   1050.0 $\pm$ 29.00    & 2, 3   \\
0657+320    &  3420   &  \phantom{1}618.0 $\pm$ 18.20  &   \phantom{1}885.0 $\pm$ 26.10  &    \phantom{1}866.0 $\pm$ 23.90  &    \phantom{1}661.0 $\pm$ 18.30    & 2, 4, 10 \\
0747+073A   &  1102B  &  \phantom{1}356.0 $\pm$ \phantom{1}9.82  &   \phantom{1}626.0 $\pm$ 17.30  &    \phantom{1}682.0 $\pm$ 18.80  &    \phantom{1}587.0 $\pm$ 16.20    & 2     \\
0747+073B   &  1102A  &  \phantom{1}553.0 $\pm$ 15.30  &   \phantom{1}848.0 $\pm$ 23.40  &    \phantom{1}866.0 $\pm$ 23.90  &    \phantom{1}673.0 $\pm$ 18.60    & 2     \\
1444$-$174  &  3866   &  \phantom{1}693.0 $\pm$ 19.20  &   \phantom{1}991.0 $\pm$ 27.40  &    \phantom{1}932.0 $\pm$ 25.80  &    \phantom{1}738.0 $\pm$ 20.40    & 2, 5   \\
1820+609    &  4054   &  1450.0 $\pm$ 40.00  &   2050.0 $\pm$ 56.70  &   2020.0 $\pm$ 55.80  &   1590.0 $\pm$ 43.80    & 6, 7   \\
2054$-$050  &  812B   &  \phantom{1}495.0 $\pm$ 13.70  &    \phantom{1}840.0 $\pm$ 23.20  &    \phantom{1}924.0 $\pm$ 25.50  &    \phantom{1}832.0 $\pm$ 23.00    & 6, 8, 9 \\

\\
\enddata
\tablenotetext{a}{Fluxes are in $10^{-18}\,$erg/s/cm$^2/$\AA \ at Earth. $UBVRI$ fluxes are on the
                  Landolt system \citep{hb06}.}
\tablenotetext{b}{Sources of photometry: 1. \citet{dahn88}, 2. \citet{brl97}, 3. \citet{liebert79}, 4. \citet{monet92}, 5. \citet{liebert88}, 6. \citet{blr01},
                  7. \citet{liebert83}, 8. \citet{eg67}, 9. \citet{dahn82}, 10. \citet{harrington93}}
\label{tab_opt_fluxes}
\end{deluxetable}
\clearpage

\begin{deluxetable}{lcccccccc}
\tabletypesize{\scriptsize}
\tablewidth{0pt}
\rotate
\tablecaption{Adopted infrared photometric fluxes\tablenotemark{a}}
\tablehead{
\colhead{WD} & \colhead{GJ} & \colhead{$J$}  &  \colhead{$H$}  &  \colhead{$K_s$}  &  \colhead{[3.6]} &  \colhead{[4.5]}  & \colhead{[5.8]} &  \colhead{[8.0]} 
}
\startdata

0230$-$144  &  3162   &    491.0 $\pm$ 13.60  &    227.0 $\pm$ 10.00  &     \phantom{1}91.8 $\pm$  5.75  &   16.870 $\pm$ 0.527  &   \phantom{1}6.9580 $\pm$ 0.227  &  2.9430 $\pm$ 0.152  &  0.865 $\pm$ 0.089 \\ 
0357+081    &  3259   &    459.0 $\pm$ 16.10  &    210.0 $\pm$ 10.90  &     \phantom{1}95.1 $\pm$  5.00  &   14.690 $\pm$ 0.462  &   \phantom{1}6.1380 $\pm$ 0.203  &  2.3340 $\pm$ 0.147  &  0.698 $\pm$ 0.069 \\ 
0657+320    &  3420   &    298.0 $\pm$ 10.70  &    155.0 $\pm$  \phantom{1}7.15  &     \phantom{1}57.7 $\pm$  4.36  &   10.150 $\pm$ 0.323  &   \phantom{1}4.2770 $\pm$ 0.145  &  1.8190 $\pm$ 0.129  &  0.631 $\pm$ 0.083 \\ 
0747+073A   &  1102B  &    298.0 $\pm$ 10.20  &    126.0 $\pm$  \phantom{1}9.31  &     \phantom{1}53.5 $\pm$  5.29  &    \phantom{1}9.550 $\pm$ 0.304  &   \phantom{1}4.0950 $\pm$ 0.140  &  1.5410 $\pm$ 0.130  &  0.501 $\pm$ 0.100 \\ 
0747+073B   &  1102A  &    316.0 $\pm$ 11.40  &    149.0 $\pm$  \phantom{1}9.19  &     \phantom{1}59.4 $\pm$  5.42  &   10.900 $\pm$ 0.345  &   \phantom{1}4.6120 $\pm$ 0.154  &  1.8390 $\pm$ 0.155  &  0.517 $\pm$ 0.076 \\ 
1444$-$174  &  3866   &    322.0 $\pm$  \phantom{1}8.59  &    160.0 $\pm$  \phantom{1}6.93  &     \phantom{1}54.6 $\pm$  5.44  &   10.970 $\pm$ 0.350  &   \phantom{1}4.6970 $\pm$ 0.158  &  2.0440 $\pm$ 0.161  &  0.783 $\pm$ 0.080 \\ 
1820+609    &  4054   &    719.0 $\pm$ 21.20  &    344.0 $\pm$  \phantom{1}9.50  &    128.0 $\pm$  6.15  &   25.860 $\pm$ 0.798  &  10.8900 $\pm$ 0.342  &  4.4100 $\pm$ 0.183  &  1.331 $\pm$ 0.099 \\ 
2054$-$050  &  812B   &    392.0 $\pm$ 29.30  &    172.0 $\pm$ 21.20  &    \phantom{1}78.8 $\pm$  9.89  &   12.780 $\pm$ 0.563  &   \phantom{1}5.2130 $\pm$ 0.404  &  0.9794 $\pm$ 0.758  &  0.384 $\pm$ 0.131 \\ 

\\
\enddata
\tablenotetext{a}{Fluxes are in $10^{-18}\,$erg/s/cm$^2/$\AA \ at Earth.  $JHK_s$ are from the 2MASS \citep{2mass}. The last four columns are {\it Spitzer} IRAC
                  fluxes \citep{kilic09a}.}
\label{tab_ir_fluxes}
\end{deluxetable}
\clearpage

\begin{deluxetable}{lcccclcc}
\tablewidth{0pt}
\tablecaption{Atmospheric parameters\tablenotemark{a} }
\tablehead{
\colhead{WD} & \colhead{$\teff$}  &  \colhead{$\log g$}   &  \colhead{$\log {\rm He/H}$}  &  \colhead{$\log {\rm He/H}$\tablenotemark{b}}  &  \colhead{Age\tablenotemark{c}}  \\
\colhead{}   & \colhead{(K)}      &  \colhead{(cm/s$^2$)} &  \colhead{}                   &  \colhead{(range)}                             & \colhead{(Gyr)} }
\startdata

0230$-$144   & ${\bf 5528\pm15}$ & ${\bf 8.15\pm0.10}$ & ${\bf\ \ -9}$       & ${\bf \le -0.3}$   &  ${\bf 4.5\pm1.0}$ \\[5pt]
0357+081     & ${\bf 5565\pm18}$ & ${\bf 8.05\pm0.12}$ & ${\bf \quad -9}$    & ${\bf \le -0.4}$   &  ${\bf 3.4\pm1.0}$ \\[5pt] 
0657+320     &      $4991\pm15$  &      $8.09\pm0.03$  &      $\quad -9$     &                    &              \\ 
             & ${\bf 4999\pm15}$ & ${\bf 8.09\pm0.03}$ & ${\bf \quad -1}$    & ${\bf \le 0}$      &  ${\bf 6.7\pm0.3}$ \\[5pt] 
0747+073A\tablenotemark{d}    &      $4496\pm51$  &     $7.97\pm0.04$  &     $\quad -9$    &    &          \\ 
             & ${\bf 4354\pm40}$ & ${\bf 7.84\pm0.04}$ & ${\bf \quad -0.3}$ & ${\bf -1.0}$ to ${\bf -0.1}$  &  ${\bf 6.3\pm0.5}$  \\[5pt] 
0747+073B\tablenotemark{d}    &      $4867\pm27$  &     $8.05\pm0.03$  &     $\quad -9$    &    &          \\ 
             & ${\bf 4723\pm37}$ & ${\bf 7.95\pm0.03}$ & ${\bf \quad \phantom{-}0}$ &  ${\bf \le  0.2}$  & ${\bf 6.3\pm0.4}$   \\[5pt] 
1444$-$174   & ${\bf 5205\pm14}$ & ${\bf 8.49\pm0.08}$ & ${\bf \quad -9}$    & ${\bf \le -0.7}$   &  ${\bf 8.1\pm0.2}$ \\[5pt] 
1820+609     &      $4907\pm11$  &      $7.96\pm0.09$  &      $\quad -9$     &                    &              \\ 
             & ${\bf 4921\pm11}$ & ${\bf 7.96\pm0.09}$ & ${\bf \quad -1}$    & ${\bf \le -0.1}$   &  ${\bf 5.6\pm1.1}$ \\[5pt] 
2054$-$050   &      $4517\pm23$  &      $7.84\pm0.09$  &      $\quad -9$     &                    &              \\ 
             & ${\bf 4491\pm42}$ & ${\bf 7.84\pm0.10}$ & ${\bf \quad -0.3}$  & ${\bf \le 0.1}$    &  ${\bf 5.9\pm1.1}$ \\ 
\\
\enddata
\tablenotetext{a}{Bold entries represent the best fit parameters for each star, i.e. those that give the lowest $\chi^2$
                  when fitting the full set of photometric data.}
\tablenotetext{b}{Range of He abundances that give equally good fits to the data within the uncertainties. See text.}
\tablenotetext{c}{White dwarf cooling age, excluding the prior phases of evolution.}
\tablenotetext{d}{Parameters based on fits that exclude the STIS data. See text.}
\label{tab_fits}
\end{deluxetable}

\clearpage

\end{document}